\documentclass[lettersize,journal]{IEEEtran}
\usepackage{amsmath,amsfonts}
\usepackage{algorithm}
\usepackage{setspace}
\usepackage{algpseudocode}
\usepackage{array}
\usepackage{textcomp}
\usepackage{stfloats}
\usepackage{url}
\usepackage{verbatim}
\usepackage{graphicx,caption}
\usepackage{epstopdf}
\usepackage{cite}
\usepackage{titlesec} %% Titles/Figures/Equations spacing
\usepackage{soul,xcolor}
\usepackage[capitalise]{cleveref}
\usepackage{amsthm}
\usepackage{amssymb}
\usepackage{makecell}
\usepackage{colortbl}
\usepackage{multirow}
\usepackage{hhline}
\usepackage{float}
\usepackage{booktabs}
\usepackage{rotating}

\newtheorem{theorem}{Theorem}

\usepackage{breqn}%% Mohamed Naguib Break Eqn
\usepackage{adjustbox}

\begin{document}
%%%%%%%%%% Mohamed Naguib Algorithm
\renewcommand{\algorithmicrequire}{\textbf{Input:}}
\renewcommand{\algorithmicensure}{\textbf{Output:}}
%%%%%%%%%%
%%%%%%%%%% Mohamed Naguib Figures
\def\img#1#2#3{
\begin{minipage}[b][.27\linewidth]{.33\linewidth}\centering
\includegraphics[width=\linewidth,height=.75\linewidth,keepaspectratio]{#1}\par
\captionof{figure}{#2}\label{#3}\end{minipage}}
%%%%%%%%%%
\title{Continuous Beam Alignment for Mobile MIMO}
\author{Mohamed~Naguib,~\IEEEmembership{Member,~IEEE,} Yahia~Shabara,~\IEEEmembership{Member,~IEEE,} \\and~Can~Emre~Koksal,~\IEEEmembership{Senior~Member,~IEEE}
\thanks{An earlier version of this paper was presented in part at the 2022 IEEE 56th Asilomar Conference on Signals, Systems, and Computers (ASILOMAR)\cite{naguibC}. (Corresponding author: Mohamed Naguib.)}
\thanks{The authors are with the Department of Electrical and Computer Engineering, The Ohio State University, Columbus, OH 43210 USA (e-mail:
naguib.9@osu.edu, shabara.1@osu.edu, koksal.2@osu.edu).

}
}

\maketitle
\begin{abstract}
Millimeter-wave transceivers use large antenna arrays to form narrow high-directional beams and overcome severe attenuation. Narrow beams require large signaling overhead to be aligned if no prior information about beam directions is available. Moreover, beams drift with time due to user mobility and may need to be realigned. Beam tracking is commonly used to keep the beams tightly coupled and eliminate the overhead associated with realignment. Hence, with periodic measurements, beams are adjusted before they lose alignment. We propose a model where the receiver adjusts beam direction “continuously” over each physical-layer sample according to a carefully calculated estimate of the continuous variation of the beams. In our approach, the change of direction is updated using the rate variation prediction of beam angles via three different solutions.
Our approach incurs no additional overhead in pilots, yet, the performance of beam tracking is improved significantly. Numerical results reveal an SNR enhancement associated with reducing the MSE of the beam directions. In addition, our approach reduces the pilot overhead by 60\% and up to 87\% while achieving a similar total tracking duration as the state-of-the-art.
\end{abstract}


\section{Introduction}
\IEEEPARstart{W}{ireless} networks of today operate at very high spectral efficiency thanks to their accumulated enhancements over the past years.
Such enhancements include the adoption of efficient waveforms \cite{hazareena2018survey} (e.g., OFDM, DFT-S-OFDM), channel coding enhancements \cite{arora2020survey} (e.g. various LDPC variants and polar codes), network densification \cite{bhushan2014network}, massive MIMO \cite{marzetta2015massive}, dynamic spectrum sharing \cite{ahmad20205g}, etc. However, further improvement of spectral efficiency will likely be marginal since we are already very close to the Shannon capacity. In order to support higher data rates, we need to venture into new spectrum bands. This effort has already been started in 5G and WiFi (802.11ad/ay).

High-frequency bands, e.g., millimeter-wave, suffer from severe attenuation \cite{Overview,OV6G}. Hence, they have not been used for cellular applications or WiFi until very recently with the advent of 5G and 802.11ad. High-frequency systems are made possible by using large antenna arrays, which can make up for the lost received signal power (due to severe path loss). In particular, a large antenna array can form highly-directional antenna beams that can focus their power/gain through narrow angular directions. By doing that, a transmit antenna can focus its transmitted energy into the most useful spatial direction(s). Similarly, a receive antenna can focus its received gain through the direction(s) that have the highest received signal power. Together, the Transmitter (TX) and Receiver (RX) antennas can provide enough gain that offsets the path loss \cite{akyildiz2018combating}.

Identifying the useful spatial directions in the channel environment and adjusting the transmit and receive antenna patterns towards such directions is commonly referred to as ``Beam Alignment''. This alignment process is challenging because as the antenna arrays get larger (to improve their gain), the beams get narrower, and the number of possible directions we need to probe for finding the ``best'' spatial directions increase as well. Also, since beam alignment needs to occur at both TX and RX, this means we need to find an optimal TX-RX beam pair. In turn, the computational complexity of beam alignment becomes quadratic in the number of beams. For instance, if the number of beams at each of TX and RX is $N$, then beam alignment requires $O(N^2)$ measurements to find the best pair. More generally, if the numbers of TX and RX beams are $N_t$ and $N_r$, respectively. Then, the complexity becomes $O(N_t \times N_r)$ measurements. Note that these complexity calculations are based on beam sweeping measurement designs and assume a simple analog transceiver. More advanced measurement designs rely on research results that show that high-frequency spectrum bands exhibit a sparse nature, where only a few significant channel paths exist. By exploiting the sparsity nature of the channel, the order of complexity figure can be shown to be $O(k^2 \log N_t/k \log Nr/k)$ \cite{beamsweep,ShabaraHow}, where $k$ is the number of channel paths. Despite the latter complexity figure being much lower, it still constitutes a large overhead that would best be minimized.

The complexity of beam alignment is an especially concerning bottleneck for communication nodes with mobility since optimal beam pairs can rapidly change. 
A repeat of the beam alignment process each time the received signal to noise (SNR) ratio drops (i.e., the beams become misaligned) entails a significant overhead. 
In fact, a repeat of the entire process may not be
required since even as the beams become misaligned, we already have prior knowledge we can exploit as to what the optimal beam pair was just a short time ago. We can exploit this prior information to narrow down the search scope of the new best beam pair. 
This process is usually referred to as ``Beam Tracking'', and it significantly reduces the number of needed measurements.

For beam tracking, a preemptive measurement strategy is commonly followed, where the transmitter sends periodic pilot symbols to allow for beam refinement. In other words, we do not wait until a beam failure event happens, but rather periodically adjust beams to keep them tightly aligned. This makes the beam tracking process smooth and keeps the communication link(s) active, without interruption. An always active communication link is indeed desirable, and if the TX sends beam tracking symbols very frequently, then it is less likely that beams will become misaligned. However, this increases the beam alignment overhead, which is not acceptable either. Hence, a tradeoff arises here where the TX needs to send pilot symbols with a frequency that is not too high so that it wastes communication resources, but also not too low that it risks beam failure.

Many beam tracking solutions have been proposed over the past few years. We can generally divide such solutions into two broad classes. The first class is where communication system resources alone are used to track the beam(s) (i.e., using in-band pilot symbols). The second class is where \textit{outside information} is either fully or partially used to track the beam(s).
This paper focuses on the first class of solutions.
To the best of our knowledge, all such solutions (of the first class) adjust beam directions only upon channel measurement. This means that between channel measurements the beams are kept fixed, and only when a pilot symbol (dedicated for tracking) is received, does the receiver update its beam, and report the measurement result to TX to allow it to update its beam, as well. 
% Thus, beams are tracked and updated in a discrete fashion. We call this tracking approach the ``Discrete'' tracking framework (or discrete tracking, or discrete framework for brevity).
Thus, beams are tracked and updated discretely. We call this tracking approach the ``Discrete'' tracking framework (or discrete tracking, or discrete framework for brevity).

In this paper, we propose an alternative framework where beams are updated at the time scale of a single (or a few) physical-layer symbol ("continuous"), even though we still rely on the usual pilot measurements ("discrete"), without any change in the underlying communication structure.
We refer to this as the ``Continuous-Discrete'' tracking framework.
The motivation behind this approach is simple: since paths continuously change directions, beam tracking can be also adapted to follow the variation at the lower time scale. This approach is in contrast to traditional "discrete" beam tracking, which updates the beam direction upon pilot observations and keeps them fixed in between. 
We show that the average received SNR can be improved substantially with the more agile, continuous tracking approach. Conversely, one can use the improvements to decrease the pilot frequency, and consequently increase the throughput.

The idea we present in this paper constitutes an umbrella or a family of beam tracking solutions that various solutions may fall under. Hence our use of the word ``framework''.
To demonstrate the benefits of our proposed framework, we will show three examples of solutions that benefit from continuous-discrete tracking when compared to their discrete tracking versions.
Two of those have their origins in the already existing literature of discrete tracking (namely, the Extended Kalman Filter (EKF) solution \cite{EKF}, and the Fast Beam Tracking (FBT) solution \cite{Fast}).
The third solution, called Main-Lobe Tracking, is newly presented in this work.
The ability to enhance existing discrete tracking solutions demonstrates a powerful aspect of our proposed idea, which is versatility.
Moreover, since our proposed framework does not necessitate any new signaling overhead (for example, no extra feedback is required, nor is it necessary to change the patterns of pilot symbols), it can readily be adopted in commercial devices with minimal to no standardization support (e.g., no impact on 3GPP or WiFi specifications).
In fact, it can be as straightforward as a firmware update for some devices. 
% The continuous-discrete approach was presented in our prior work\cite{naguibC}, while the full contribution in this paper can be summarized as:
% Our contributions in this paper can be summarized as:
{The continuous-discrete approach was presented in our prior work \cite{naguibC}. The full contribution in this paper can be summarized as follows:}
\begin{itemize}
    \item We propose a framework for beam tracking, featuring continuous beam updates.
    \item We illustrate that continuous, smooth beam adjustments, accompanying discrete jumps at measurement instances, yield major power (SNR) gains. As a result, at a given desired SNR, the frequency of pilots can be reduced significantly.
    \item We provide three simple tracking algorithms, under our framework, based on different techniques.
    \item We provide an analysis of different antenna arrangements, including linear and planar structures, and study their respective robustness to variations in the beam direction.
    \item We show that due to shortened beam coherence times associated with narrower beams, larger MIMO arrays do not necessarily lead to improved performance, as they require more frequent beam updates in analog beamforming.
\end{itemize}

Next, we further discuss the related work.
In\cite{Fast}, a Fast Beam Tracking (FBT$_{D}$) approach was derived to make the MSE converge faster to the Cramér-Rao lower bound (CRLB). In this approach, both the initial beam discovery and beam tracking problems are considered. A coarse beam sweeping is proposed to discover the initial beam direction, then a recursive beam tracking algorithm is proposed to minimize the MSE of the true Angle of Arrival (AoA).
In beam tracking measurements, noise is added to the AoA/AoD in a non-linear fashion at the antenna elements. This makes the EKF (a non-linear version of the Kalman Filter) a suitable solution for this problem. In \cite{EKF}, an Extended Kalman Filter with discrete beam updates, i.e., EKF$_{D}$, is proposed. One of the solutions we present in this paper follows a similar approach except that the state transition model is formulated as a continuous one.
In \cite{PF}, an Auxiliary Particle Filter is proposed as another tracking algorithm that deals with the nonlinear noisy function of the AoA. In this approach, a group of particles is generated at each channel measurement. 
An estimate of the AoA is found by averaging the estimate of all generated particles. 
In addition, the beam tracking solutions adopted in 5G \cite{5GNewRadio} and WiFi \cite{cordeiro2010ieee} both rely on the same beam refinement procedure followed in the initial beam discovery process, e.g., using narrow CSI-RS beams in 5G.

Recall that a second class of tracking solutions exists which do not rely on measurements of the wireless communication channel. Such solutions rely on side information to estimate the position of the other node (either the TX or RX) and consequently point its beam toward it. Such solutions may also be capable of predicting beam blockages and thus can react accordingly \cite{CVABT}. Examples of such solutions can be radar imaging, LIDAR \cite{klautau2019lidar}, GPS, and cameras. For instance, in \cite{chou2021fast} the GPS location of one device is shared with the other node as helping information to assist in beam selection. In \cite{jiang2022computer}, a camera is used to acquire a sequence of images which are fed into a Machine Learning model that is used to predict future beams.

\textbf{Paper Organization:} We provide an illustrative example of Continuous-Discrete tracking in Section \ref{sec:motivation}. The system model is proposed in Section \ref{sec:model}, and Section \ref{sec:approach} demonstrates the Continuous-Discrete beam tracking approaches. Section \ref{sec:LRT} provides an analytical discussion of a complementary problem to the beam tracking which is how to choose the pilot period. Section \ref{sec:Nres} shows simulation results comparing the Continuous-Discrete proposal to its counterparts and an evaluation of methods of selecting the Pilot Period. Finally, the conclusion is presented in Section \ref{sec:conc}.

\section{Illustrative Example}\label{sec:motivation}
In this section, we provide a simplified example to demonstrate our proposed Continuous-Discrete beam tracking solution, and compare it against the traditional discrete beam tracking approaches. The fusion of discrete measurements and continuous AoA updates (which enables continuous beam updates) is referred to as "Continuous-Discrete" Beam Tracking.

\begin{figure}
    \centering
    \includegraphics[scale=0.45]{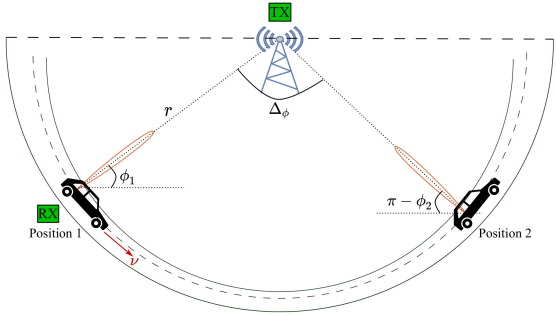}
    \caption{Example of a moving receiver with a Continuous-Discrete tracking}\label{fig:motExp}
\end{figure}

In the example shown in Fig. \ref{fig:motExp}, a TX with a single antenna serves a moving RX with a directional antenna gain. 
The RX moves on along the shown circular path with a constant velocity $\nu$. Let us assume that at position 1 the RX has accurately discovered 
the AoA, $\phi$, of the strongest channel path, and has aligned its receive antenna beam towards that path. After that, the RX keeps tracking $\phi$ using periodic pilot symbols that arrive every $T$ seconds. 
% Upon the arrival of a pilot symbol, 
Upon a pilot symbol arrival, 
the RX updates the estimate of the AoA, $\phi$ and its slope variation, $\frac{d\phi}{dt}$. Using the slope variation, the RX predicts the change in the AoA, and continuously updates the beamforming vectors to stay tightly coupled with the incoming path.

Recall that RX moves with constant angular velocity. Hence, intuitively, as the prediction errors of the AoA and the slope variations go to zero, the period $T\to \infty$, which means that pilot symbols are not needed to keep tracking the AoA. That is because $\frac{d\phi}{dt}$ perfectly tells us exactly what the value of $\phi$ will be at any future time.
On contrary, "Discrete" Beam Tracking solutions only update the AoA \textbf{once} every pilot symbol arrival. That is, discrete tracking does not update the AoA estimate in the time duration in between pilot arrivals. Also, discrete tracking does not rely on AoA rate of change information.
Hence, the misalignment between the beam direction and the incoming paths increases as the pilot period increase. This means that, in a Discrete tracking approach the RX needs to minimize the period $T$ (increase the pilot overhead) as the beamwidth decreases (array size increase). 
Our proposed Continuous-Discrete measurement framework incorporates extra system information embodied in the rate of change of the AoA.
This extra information allows us to predict how the AoA changes even when no measurements are available. This, in turn, allows us to extend the pilot period without significantly degrading the link quality nor risking losing beam alignment.

\begin{figure}
    \centering
    \includegraphics[scale=0.48]{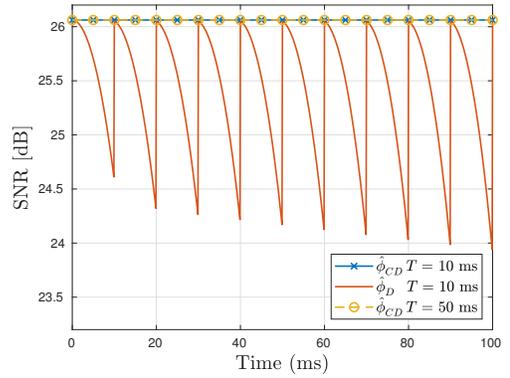}
    \caption{SNR for Continuous-Discrete and Discrete tracking}\label{fig:motExpres}
\end{figure}
Now, let us introduce how the AoA changes based on the special scenario assumed in Fig. \ref{fig:motExp}. The RX at position 1 is moving with constant velocity $\nu$, and the AoA changes continuously based on the following equation:
\begin{equation}
    \begin{aligned}
    \phi(t) = \phi^{}_0 + \frac{\nu}{r}\,t,
    \end{aligned}
\end{equation}
\noindent where $r$ is the distance from the TX to the RX. So, after a pilot period $T$, the RX will move to position 2, and the new AoA is $\phi^{}_{2} = \phi^{}_{1} + \frac{\nu}{r}\,T$. The Continuous-Discrete tracking allows us to update the beam direction continuously based on a prediction of the angular velocity $\frac{\nu}{r}$, while Discrete tracking only updates the beam direction upon pilot arrival, (i.e., position 1,2). Fig. \ref{fig:motExpres}, shows the SNR for the example shown in Fig. \ref{fig:motExp}, under both Discrete and Continuous-Discrete measurement frameworks. For that experiment, the total tracking time is 100 ms, $\nu = 100$ Km/hr, $r = 20$ m, and $\phi^{}_{0} = \pi/4$, and the two values for the pilot period $T = \{10,50\}$ ms. The results from Fig. \ref{fig:motExpres} reveal the weakness of the Discrete tracking approaches, where the instantaneous SNR drops drastically in between pilot symbols. To overcome the drop in the SNR, the pilot period should be decreased, i.e., $T<10$ ms. On the other hand, the prediction of the rate of change of the AoA makes the Continuous-Discrete framework to stay tightly coupled even when only two pilot symbols are used over the entire tracking duration (i.e., $T=50$ ms). Furthermore, the Continuous-Discrete approach can stay aligned without any pilot symbols as long as the prediction of the slope variation is available. 

In this example, we have considered an oversimplified scenario. In a real system, however, the channel environment becomes more complex, requiring more sophisticated mathematical modeling and tracking techniques. Modeling real system complexities will be tackled in the following sections starting with the System Model in Section \ref{sec:model}.
We will see that errors in the prediction of the slope variation force the Continuous-Discrete solution to limit the pilot period to alleviate these errors. Yet, Continuous-Discrete approaches still can do better than the Discrete approach, as measured in two metrics: (1) Low Pilot Overhead, where the pilot period is longer than the period of the Discrete case. (2) SNR, where the Continuous-Discrete can have higher SNR than the Discrete case for the same pilot period. 
\section{System Model}\label{sec:model}
\subsection{Notation}
The lowercase $x$ and $\boldsymbol{x}$ denote scalar and vector quantities respectively, while the uppercase $\boldsymbol{X}$ denotes a matrix.
The continuous variation in $\boldsymbol{x}$ is denoted by $\boldsymbol{{x}}(t)$, and $\boldsymbol{{x}}^{}_k$ is the sampled version of  $\boldsymbol{{x}}(t)$ each $T$ seconds, where $\boldsymbol{{x}}^{}_k \triangleq \boldsymbol{{x}}(t)\,|_{\,t=k\,T}$. 
The first order derivative of $\boldsymbol{x}(t)$ is $\boldsymbol{\dot{x}}(t) = \frac{d \boldsymbol{x}(t)}{dt}$. Moreover, $(.)^{\dag}$, $\Re{(.)}$, and $\Im{(.)}$ denote the conjugate transpose, the real and the imaginary parts respectively. Finally, The standard Q function is denoted by $\mathbb{Q}$, while its inverse is denoted by $\mathbb{Q}^{-1}$.
\subsection{Channel Model}
We consider a Single Input Multiple Output (SIMO) channel model with a single antenna at the transmitter, and $N$ antennas at the receiver. The receiver antennas are equally spaced and arranged linearly to form a Uniform Linear Array (ULA).
We assume an analog beamforming transceiver architecture, where a single RF chain is utilized, and a phase shifter is applied for each antenna element in the array. We assume a 2D multi-path channel model from \cite{HeatChmodel} with a channel impulse response:
\begin{equation}\label{eqn:chn}
        \boldsymbol{c}(t) = \sum_{\ell=1}^{L}\alpha^{}_{\ell}(t)\,e^{j2 \pi f^{}_{d,\ell}t} \boldsymbol{a}^{}_{R}\left({\phi^{}_{\ell}(t)}\right),
\end{equation}
\noindent where $\ell$ is the path index, $L$ is the total number of paths, $\alpha^{}_{\ell}(t)$ is the complex path gain, $f^{}_{d,\ell}$ is the Doppler shift, $\boldsymbol{a}^{}_{R}$ is the array steering vector, and $\phi^{}_{\ell}(t)$ is the time-varying AoA, which changes over time due to mobility of the transmitter, receiver and the environment. The steering vector is given by:
\begin{equation}
\boldsymbol{a}^{}_{R}(\phi) =\begin{bmatrix}
1\,\,e^{-j2\pi \Delta \cos\left(\phi\right)}\,\,
\cdot\cdot\cdot\,e^{-j2\pi \Delta \left(N-1\right) \cos\left(\phi\right)}
\end{bmatrix}^T,
\end{equation}
\noindent where $\phi \in [0\,,2\pi)$, $\Delta$ is the antenna spacing normalized by the wavelength, and we assume $\Delta=0.5$ for the rest for this paper, and $N$ is the number of array elements. The variation in the AoA is given by:
\begin{equation}\label{eqn:phit}
\begin{aligned}
    \phi^{}_{\ell}(t) &= \phi^{}_{0,\ell} + \int_0^t{\dot{\phi}^{}_{\ell}(\tau)\,d\tau}\\
    \dot{\phi}^{}_{\ell}(t) &= {\dot{{\phi}}^{}_{i,\ell}} + \mathcal{B}^{}_{\ell}(t),
\end{aligned}
\end{equation}
\noindent where $\phi^{}_{0,\ell}$, and ${\dot{{\phi}}^{}_{i,\ell}}$ are the initial AoA and the initial rate of change for the AoA, respectively. The variation term $\dot{\phi}^{}_{\ell}(t)$ denotes the rate of change in AoA.
We model the changes in $\phi(t)$ through modeling its rate of change $\dot{\phi}(t)$. In a general environment, users can accelerate, decelerate, rotate and may move on curved paths. 

The objective of this paper is not to provide detailed models of motion and integrate them into a cumbersome calculus. Instead, we aim to convey the value of incorporating the variations into beamforming. As a result, we use an approach in which the vagaries of the mobility are summarized in a general stochastic model. In particular, as a representation of $\dot{\phi}(t)$, we use $\mathcal{B}^{}_{\ell}(t)$, a zero mean Brownian motion process with variance $Q\,t$, where $d\mathcal{B}^{}_{\ell}(t) = \omega^{}_{\ell}(t)\,dt$, and $\omega^{}_{\ell}(t)$ is a zero-mean White Gaussian process with a Power Spectral Density (PSD) $Q$ for all frequencies $-\infty < f < \infty$. 
A small value of Q means that $\dot{\phi}(t)$ is almost fixed (and that changes in $\phi(t)$ is steady), and vice versa for large values of Q.

Given the sparse nature of the mmWave channel, the received power from all directions is limited to a few sharply defined AoAs \cite{mmwavesparsity1,mmwavesparsity2}. In addition, since our receiver has an analog architecture, it can only form one directional beam towards the strongest path in the channel (i.e., the path with the highest received power). Hence, we can assume that all other paths are highly attenuated since they are weaker and lie outside the main lobe of the receive antenna. Therefore, we omit the subscript $\ell$ for paths and treat attenuated weaker paths as noise.
\begin{figure}[t]
 \centering
\includegraphics[scale=0.55]{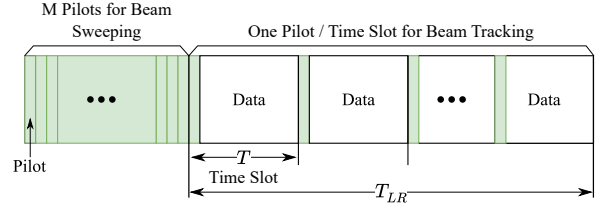}
\caption{Frame structure of length $T^{}_{LR}$\cite{Fast}}\label{fig:Frame}
\end{figure}

{\bf Communication Protocol:} Using the frame structure in Fig. \ref{fig:Frame}, we have two stages for beam alignment: the first stage consists of $M$ pilots required for \textit{initial beam alignment}, and the second stage applies one pilot each time slot for \textit{beam tracking}. The second stage can be viewed as a time frame consisting of $k$ time slots. Each time slot has a duration $T$ seconds, while the total duration of the time frame is $T^{}_{LR}$ seconds (i.e., $T^{}_{LR}=k\times T$).

In this paper, we only focus on the second stage that is concerned with beam tracking. Hence, the initial beam alignment performed at the first stage is out of the scope of this work, however, we assume that we have a good initial beam direction. The $k^{th}$ received pilot sample is given by:
\begin{equation}\label{measurement}
    \begin{aligned}
        z^{}_k &= \alpha\,p\,e^{j2\pi f^{}_{d} k T}\,\boldsymbol{w}^\dag(\bar{\phi}^{}_k) \, \boldsymbol{a}^{}_{R}\left({\phi^{}_k}\right) + v^{}_k\\
        &= \displaystyle\frac{\alpha\,p\,e^{j2\pi f^{}_{d} kT}}{\sqrt{N}}\,\sum_{m = 0}^{N-1} e^{-j2\pi\Delta m[\cos({\phi^{}_k}) - \cos({\bar{\phi}^{}_k})]}+ v^{}_k,
    \end{aligned}
\end{equation}
% \noindent where {$\alpha$ is considered a fixed-known value in this context due to the assumption that the complex path gain changes at a much slower rate than the AoA during a frame time. This assumption allows for a simplified model where $\alpha$ can be treated as a constant parameter throughout the analysis as considered in \cite{Fast}.}
\noindent where {$\alpha$ is considered a fixed-known value since the scope of the paper is beam-tracking, in our model, we assume that the complex path gain ($\alpha$) changes at a much slower rate than the AoA during the frame time. To us, the important point is not that beam-tracking has less of a value with slow angular variations relative to the path gain. Instead, we claim that under a wide variety of situations, the continuous-discrete approach performs much more favorably compared to the traditional purely discrete approach. This assumption allows for a simplified model where $\alpha$ can be treated as a constant parameter throughout the analysis as considered in \cite{Fast}.}
The amplitude for a known pilot symbol is denoted by $p$, $v^{}_k \sim \mathbb{C}\mathcal{N}(0,\sigma^2_{v})$ is a circularly symmetric complex Gaussian noise sample, $\bar{\phi}^{}_k$ is the beam pointing direction, and $\boldsymbol{w}(\bar{\phi}^{}_k)$ is the beamforming vector which is given by:
\begin{equation}\label{eqn:bmf}
    \boldsymbol{w}(\bar{\phi}^{}_k) {=} \frac{1}{\sqrt{N}}\hspace{-3px}\begin{bmatrix}
1\,\,e^{-j2\pi \Delta \cos(\bar{\phi}^{}_k)}\hspace{-1px}\,
\hspace{-1px}\cdot\cdot\cdot\,e^{-j2\pi \Delta (N-1) \cos(\bar{\phi}^{}_k)}
\end{bmatrix}^T\hspace{-10px}
\end{equation}
Succinctly, we can rewrite \eqref{measurement} normalized as follows:
\begin{equation}\label{meas}
    y^{}_k = \displaystyle\frac{1}{N}\sum_{m = 0}^{N-1}e^{-j2\pi\Delta m[\cos({\phi^{}_k}) - \cos({\bar{\phi}^{}_k})]} + \displaystyle\frac{n^{}_k}{\sqrt{\rho}},
\end{equation}
\noindent where $n^{}_k \sim \mathbb{C}\mathcal{N}(0,1)$, $\rho = N\,{|\alpha\,p|^2}/{\sigma^2_{v}}$ is the SNR including the array gain, and ${|\alpha\,p|^2}/{\sigma^2_{v}}$ is SNR for each antenna element. For more straightforward dealing with \cref{meas} in the following sections, we define the signal part by:
\begin{equation}
    h\left(\phi^{}_{k},\bar{\phi}^{}_{k}\right) \triangleq \displaystyle\frac{1}{N}\sum_{m = 0}^{N-1}e^{-j2\pi\Delta m[\cos({\phi^{}_k}) - \cos({\bar{\phi}^{}_k})]}
    % \vspace{-10pt}
\end{equation}
\subsection{Uniform Planar Arrays}
Up to this point, development in \cref{eqn:chn,measurement,eqn:bmf,meas} is based on ULAs. Here, we turn our attention to Uniform Planar Arrays (UPA)  since they provide higher gain coupled with resilience to channel variations.
We assume a square UPA where $N$ elements are equally spaced and arranged over the area of a square. The steering and the beamforming vectors are given by:
\begin{equation}\label{eqn:arupa}
    \displaystyle\boldsymbol{a}^{}_{P}(\phi,\theta) = \boldsymbol{a}^{}_{R}(\theta)\otimes\boldsymbol{a}^{}_{R}(\phi),
    % \vspace{-8pt}
\end{equation}
\noindent where $\otimes$ is the Kronecker product, while $\phi$ and $\theta$ are the azimuth and elevation angles, respectively. The beamforming vector is given by $\boldsymbol{w}^{}_{P}(\bar{\phi}^{}_k,\bar{\theta}^{}_k) = \boldsymbol{a}^{}_{P}(\bar{\phi}^{}_k,\bar{\theta}^{}_k) / \sqrt{N}$. 
Now, the model for the UPA is based on \eqref{eqn:arupa}, and $\boldsymbol{w}^{}_{P}(\bar{\phi}^{}_k,\bar{\theta}^{}_k)$. Finally, if we assume AoA variation only in the azimuth direction, and we fix $\theta=90^{\circ}$, then the beamforming vector will be reduced to $\boldsymbol{w}^{}_{P}(\bar{\phi}^{}_k) = \boldsymbol{a}^{}_{P}(\bar{\phi}^{}_k) / \sqrt{N}$.

\subsection{Problem Statement}
Our objective is to minimize the instantaneous MSE between the true $\text{AoA}$ $\phi(t)$ and its estimate $\hat{\phi}(t)$, which can be stated by:
\begin{equation}
    \underset{\hat{\phi}(t)}{\min}\, \mathbb{E}[(\phi(t)-\hat{\phi}(t))^2]
\end{equation}
While traditional solutions update $\hat{\phi}(t)$ at discrete measurement instances only, which create larger MSE as $\phi(t)$ drifts in-between pilots, we update $\hat{\phi}(t)$ continuously using an estimate of the $\text{AoA}$ rate of change $\dot{\phi}(t)$ (recall equation \eqref{eqn:phit}). We define the system state as $\boldsymbol{x}(t) \triangleq \begin{bmatrix}
\phi(t)&\dot{\phi}(t) \end{bmatrix}^T$, and let $\hat{\boldsymbol{x}}(t)$ be its estimate, whose error covariance matrix is $\boldsymbol{P}(t)$, where
\begin{equation}
    \begin{aligned}
     \boldsymbol{P}(t)&\overset{\Delta}=\mathbb{E}\left[(\boldsymbol{x}(t) - \hat{\boldsymbol{x}}(t))(\boldsymbol{x}(t) - \hat{\boldsymbol{x}}(t))^T\right]\\
    \end{aligned}
\end{equation}
The estimate of $\boldsymbol{x}(t)$ which incorporates both the $\text{AoA}$ and its rate of change is tackled in the following section.
\section{Continuous-Discrete Beam Tracking Approaches }\label{sec:approach}
In traditional beam tracking, the estimate of AoA is updated using the pilot symbol that arrives each time slot, and the RX fixes the beam direction to that estimate until the arrival of the following pilot symbol. In contrast, we propose a Continuous-Discrete beam tracking approach that exploits the prediction of the rate of change of the channel variation, which is neglected in traditional approaches. The proposed approach continuously updates the beam direction to minimize the instantaneous MSE. Furthermore, the approach presented here hinges on updating the estimate of the AoA not only based on pilot symbols but also continuous predictions of the rate of variations of the AoA in between pilots. In other words, we perform continuous beam tracking instead of abrupt adjustments at each pilot.

In this section, we apply the Continuous-Discrete approach to two different techniques to prove that the proposed approach improves the MSE over discrete baseline solutions. Moreover, a novel beam-tracking algorithm that utilizes an intelligent formula to quickly find the AoA estimate with low complexity is provided. The three frameworks are listed as follows:
\begin{itemize}
    \item An Extended Kalman Filter:
     An extended version of the well-known Kalman Filter, which is more suitable to a non-linear system.
    \item A Fast Beam Tracking:
    The proposed framework in \cite{Fast}, targets converging the MSE faster to the CRLB.
    \item Main-Lobe Approximation:
    A novel approach used to track the beam direction based on an approximation of the amplitude of main-lobe proposed in Section \ref{DMLT}.
\end{itemize}
Following this, we provide a detailed discussion of the three proposed approaches.

\subsection{Approach 1: Extended Kalman Filter}
The Kalman Filter is a popular tool used in tracking problems, which motivates its use in beam tracking. However, recall from \cref{meas} that the measurement $y^{}_{k}$ is a non-linear function of the AoA ($\phi^{}_{k}$). This makes the traditional Kalman Filter unsuitable for this system.
Instead, the EKF is the non-linear version of the KF, which is more suitable for the considered non-linear system in this paper. The traditional discrete form of the EKF is presented with two equations: (1) \textbf{State}, and (2) \textbf{Measurement/Equation}. The two equations are given correspondingly as follows:
\begin{equation}\label{eqn:state_new}
    \boldsymbol{x}^{}_{k}=\mathfrak{F}(\boldsymbol{x}^{}_{k-1})+\omega^{}_{k}
\end{equation}
\begin{equation}\label{eqn:obs_new}
    z^{}_k = \mathfrak{H}(\boldsymbol{x}^{}_{k})+ v^{}_{k},
\end{equation}
\noindent where, $\mathfrak{F}(\boldsymbol{x}^{}_{k-1})$, and $\mathfrak{H}(\boldsymbol{x}^{}_{k})$ are non-linear functions of the current and previous samples of the state vector $\boldsymbol{x}^{}_{k}$. The beam tracking approach in \cite{EKF}, replaces \cref{eqn:state_new} by their discrete model of the AoA variation. On the other hand, we represent the evolution in the state equation by the continuous time variation of the AoA given by \cref{eqn:phit} as follows: 
\begin{equation}
    \begin{aligned}
        \boldsymbol{\dot{x}}(t) 
        &=\boldsymbol{A}\,\boldsymbol{{x}}(t) + \boldsymbol{g}\, \omega(t),
    \end{aligned}
\end{equation}
\noindent where, $\boldsymbol{A}=\bigl[\begin{smallmatrix} 0&1\\0&0\end{smallmatrix}\bigr]$, and $\boldsymbol{g}=\bigl[\begin{smallmatrix} 0\\1\end{smallmatrix}\bigr]$ are matrices with proper dimensions. Moreover, the measurement equation is presented by samples taken upon pilot symbol arrival which is given by \cref{meas}. Now, we have an EKF with a continuous time state model and discrete samples of observations which formulate the Continuous-Discrete Extended Kalman Filter (EKF$_{CD}$).
The derivation of the state dynamics of the EKF$_{CD}$ can be found in \cite{lewis}, but we only provide a sketch as follows: The main idea of the EKF derivation is the linearization of the signal part ($h(\boldsymbol{x}^{}_{k})$) of \cref{meas}. The following procedures are the same as that of the standard Kalman filter which can be found in \cite{lewis}.
The standard recursion equations (see \cite{lewis}) are divided into two stages: (1) \textbf{Prediction}, and (2) \textbf{Update}. The prediction stage is given by:
\begin{equation}\label{predctx}
    \hat{\dot{\boldsymbol{x}}}(t) = \boldsymbol{A}\,{\hat{\boldsymbol{x}}(t)}\\
\end{equation}
\begin{equation}\label{predctP}
    \hat{\dot{\boldsymbol{P}}}(t) = \boldsymbol{A}\,\boldsymbol{P}(t) + \boldsymbol{P}(t)\,\boldsymbol{A}^T + \boldsymbol{g}\,Q\, \boldsymbol{g}^T,
\end{equation}
\noindent where $\hat{\dot{\boldsymbol{x}}}(t)$ is the prediction of the time derivative of the state vector $\boldsymbol{x}(t)$, $\hat{\dot{\boldsymbol{P}}}(t)$ is the prediction of the time derivative of ${\boldsymbol{P}(t)}$, and $Q\,\delta(\tau) = \mathbb{E}[w(t)\,w(t+\tau)]$ is the process covariance that drives the state vector, and is assumed to be bounded by $Q$. The estimate ${\hat{\boldsymbol{x}}}(t)$ during the prediction stage is the solution of the differential equation in \eqref{predctx}. 
The update stage, which amends the $k^{th}$ iteration of the standard Kalman gain ($\mathbb{K}^{}_k$), is given by:
\begin{equation}\label{KGain}
\mathbb{K}^{}_k = \boldsymbol{P}^{k-1}_{k}\,\boldsymbol{H}_k^\dag \left(\boldsymbol{H}^{}_k\,\boldsymbol{P}^{k-1}_{k}\,\boldsymbol{H}_k^\dag +\frac{1}{\rho}\right)^{-1}
\end{equation}
\begin{equation}\label{PKu}
\boldsymbol{P}^{k}_{k} = (\mathrm{I} - \mathbb{K}^{}_k \, \boldsymbol{H}^{}_k)\boldsymbol{P}^{k-1}_{k}
\end{equation}
\begin{equation}\label{eqn:xhat}
\hat{\boldsymbol{x}}^{k}_{k} = \hat{\boldsymbol{x}}^{k-1}_{k} + \mathbb{K}^{}_k \left(y^{}_k - h(\hat{\boldsymbol{x}}^{k-1}_{k})\right),
\end{equation}
\noindent where $\hat{\boldsymbol{x}}^{k-1}_{k}$ and $\boldsymbol{P}^{k-1}_{k}$ are the estimate and the error covariance matrix at $k^{th}$ iteration given $k{-}1$ measurements.
Finally, $\boldsymbol{H}^{}_k {=} \nabla^{}_{{\boldsymbol{x}}^{}_{k}}h|_{\boldsymbol{x}^{}_k = \hat{\boldsymbol{x}}^{k-1}_{k}}$ is the signal part ($h(\boldsymbol{x}^{}_{k})$) gradient estimated at $\hat{\boldsymbol{x}}^{k-1}_{k}$.

\begin{figure}[t]
 \centering
 \includegraphics[scale=0.4]{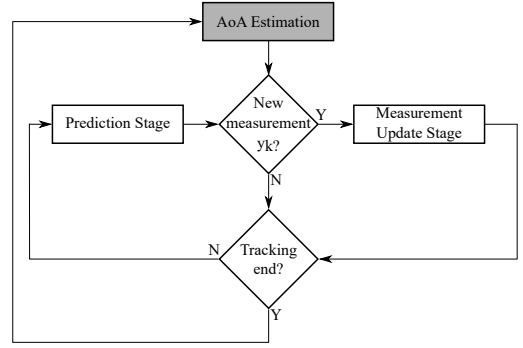}
 \caption{An overview of Continuous-Discrete EKF procedure.}\label{fig:EKFCD}
\end{figure}

An overview of the EKF$_{CD}$ procedure can be described in Fig. \ref{fig:EKFCD}. First, we assume knowledge of the estimate of the initial AoA $\phi^{}_0$. Then, the state is updated based on \cref{eqn:xhat} if a measurement is available. Otherwise,  we repeat the prediction stage $n^{}_s$ times, i.e., we make a prediction every ${T\,/\,n^{}_s}$ seconds. {Both stages complement each other, as the estimates $\hat{\boldsymbol{x}}^{k}_{k}$ and $\boldsymbol{P}^{k}_{k}$ from the update stage are utilized to initialize the prediction stage. Similarly, the final instantaneous prediction, $\hat{\boldsymbol{x}}(t)$ and $\hat{\boldsymbol{P}}(t)$, obtained in the prediction stage, serves as the initialization for the subsequent update stage.}
When the tracking period is over, we start searching for a fresh estimate of the AoA, similar to estimating $\phi^{}_0$.
In this paper, our main focus is \textit{beam tracking}, and our solution complements any \textit{beam discovery} (initial AoA estimation). Hence the latter is not included in this work. Further, optimizing the tracking period length is left as future work. {Moreover, EKF can be readily extended to incorporate the complex path gain ($\alpha$) in the state vector $\boldsymbol{x}$ for estimation upon the arrival of pilot symbols, as demonstrated in \cite{EKF}. However, our paper concentrates on AoA tracking to distinguish between discrete and continuous-discrete approaches.}

In the implementation of EKF$_{CD}$, we replaced $\boldsymbol{H}^{}_k,\,y^{}_k,$ and $h$ by $\Tilde{\boldsymbol{H}}^{}_k = \left[\Re{[\boldsymbol{H}^{}_k],\Im{[\boldsymbol{H}^{}_k]}}\right]^T,\,\Tilde{\boldsymbol{y}}^{}_k = \left[\Re{[y^{}_k],\Im{[y^{}_k]}}\right]^T,$ and $\Tilde{\boldsymbol{h}} = \left[\Re{[h],\Im{[h]}}\right]^T$, respectively, to cast the problem as a real-valued problem, similar to \cite{EKF}.
However, unlike \cite{EKF}, we utilize the pointing direction of the beamforming vector to be the previous estimate, i.e., $\bar{\phi}^{}_0 = \phi^{}_0$, and $\bar{\phi}^{}_k = \hat{\phi}_k^{k-1}$.

For UPA arrays, we extend the state vector $\boldsymbol{x}(t) \triangleq \begin{bmatrix} \phi(t)&\dot{\phi}(t)&{\theta}(t)&\dot{\theta}(t) \end{bmatrix}^T$ to include the variation in the elevation plane as well. In that case, \cref{predctx,predctP,KGain,PKu,eqn:xhat} will be a function of both the azimuth and elevation variations.
\subsection{Approach 2: Fast Beam Tracking}
The FBT$_D$ was proposed in \cite{Fast} as a low complexity algorithm that makes the MSE of the true AoA converge faster to the CRLB. In \cite{Fast}, they considered both the beam discovery and the beam tracking problems. The AoA of the {\it Discrete Fast Beam Tracking} (FBT$_{D}^{}$) is updated based on the following equation:
\begin{equation}\label{eqn:FBT}
\hat{\beta}^{}_k = \left[\hat{\beta}^{}_{k-1}- a^{}_n\,\Im{[y^{}_k]}\right]_{-1}^1,
\end{equation}
\noindent where $\hat{\beta}^{}_k = \sin(\hat{\phi}^{}_k)$, $\hat{\beta}^{}_{k-1}$ is the current and the previous estimates of $\sin({\phi}^{}_k)$ and $\sin({\phi}^{}_{k-1})$ respectively, and $a^{}_n$ is the step size. The estimate $\hat{\phi}^{}_k$ is found by taking the inverse of the sine function.
In this section, we are only concerned with extending the discrete FBT$_{D}$ (beam tracking only) to a Continuous-Discrete Fast Beam Tracking (FBT$_{CD}$) solution. 
The key idea of the Continuous-Discrete is to predict the slope variation of the AoA, so we need a prediction of the slope variation besides the estimate of the AoA itself using FBT$_D$. We take advantage of the prediction stage in the EKF$_{CD}$ to continuously predict the AoA and its slope between two pilot symbols. Two stages describe the recursion of the FBT$_{CD}$: (1) the \textbf{prediction} stage using equation \eqref{predctx}, which continuously predicts the estimate of the AoA and its rate of change found in the update stage. 
This stage should be initialized by an estimate of AoA and the slope variation estimate at the beginning of the prediction period.
(2) the \textbf{update} stage is responsible for finding the new estimates for the AoA and its slope variation, and it is accomplished through the following two steps:
\subsubsection{Discrete Fast Beam Tracking}
The first step is to update the AoA using the FBT$_{D}$ \cite{Fast}, which is given by \cref{eqn:FBT}. However, the measurement samples in our model (i.e., \cref{meas}) is normalized over the total SNR including the array gain, which is different to the model in \cite{Fast}.
To accommodate for differences from the model of \cite{Fast}, we recalculate the step size as follows:
\begin{equation}
    % a^{}_n {=} \displaystyle{\frac{\partial\log{P(y^{}_k|\hat{\beta}^{}_{k-1}\,,\boldsymbol{w}(\hat{\phi}^{}_{k-1}))}}{\partial\hat{\beta}^{}_{k-1}}}\bigg/{I\left(\hat{\beta}^{}_{k-1},\boldsymbol{w}(\hat{\phi}^{}_{k-1})\right)}
    a^{}_n = \displaystyle{\frac{\partial\log{P(y^{}_k|\hat{\beta}^{}_{k-1}\,,\boldsymbol{w}(\hat{\phi}^{}_{k-1}))}}{\partial\hat{\beta}^{}_{k-1}}}\,\bigg/\,{I^{}_{k-1}}
\end{equation}
It is easy to find the numerator and the denominator (Fisher Information) of $a^{}_n$ by the following:
\begin{equation}
    \begin{aligned}
    % I\left(\hat{\beta}^{}_{k-1},\boldsymbol{w}(\hat{\phi}^{}_{k-1})\right)&=\mathbb{E}\left[\displaystyle -\frac{\partial^2\log{P(y^{}_k|\hat{\beta}^{}_{k-1}\,,\boldsymbol{w}(\hat{\phi}^{}_{k-1}))}}{\partial^2\hat{\beta}^{}_{k-1}}\right]\\
    {I^{}_{k-1}}&\triangleq\mathbb{E}\left[\displaystyle -\frac{\partial^2\log{P(y^{}_k|\hat{\beta}^{}_{k-1}\,,\boldsymbol{w}(\hat{\phi}^{}_{k-1}))}}{\partial^2\hat{\beta}^{}_{k-1}}\right]\\
    &={2(N-1)^2\pi^2\Delta^2\rho}\\
    \end{aligned}
\end{equation}
Also,
\begin{equation}
    \frac{\partial\log{P(y^{}_k|\hat{\beta}^{}_{k-1}\,,\boldsymbol{w}(\hat{\phi}^{}_{k-1}))}}{\partial\hat{\beta}^{}_{k-1}} = 2\pi\Delta(N-1)\rho
\end{equation}
Hence, we have the step size as:
\begin{equation}
    a^{}_n = \left[{(N-1)\,\pi\,\Delta}\right]^{-1}
\end{equation}
\subsubsection{Slope Variation Update}
The FBT$_{D}$ {does not} consider updating the slope variation of the AoA ($\dot{\phi}$), we utilize the EKF$_{CD}$ in the background of FBT$_{CD}$ to update $\dot{\phi}$, which is needed for the prediction stage.
\subsection{Approach 3: Main-Lobe Tracking Algorithm}
In EKF$_{CD}$, and FBT$_{CD}$, the slope variation estimate was found using an EKF framework. Here, we introduce a Continuous-Discrete Main-Lobe Tracking Algorithm (ML$_{CD}$) based on the proposed Algorithm \ref{alg1} to update the estimate of the AoA. The estimate of the rate of change ($\hat{\dot{\phi}}$) is updated using an MMSE of the difference between two consecutive slope variation instants ($\dot{{\phi}}^{}_k - \hat{\dot{\phi}}^{}_{k-1}$). The proposed approach not only proves the validity of the Continuous-Discrete over all discrete baseline but also it provides a tractable analysis needed for the upcoming Section \ref{sec:LRT}. The ML$_{CD}$ is described by two stages as in the EKF$_{CD}$, and FBT$_{CD}$; the \textbf{prediction} stage using \cref{predctx}, and \textbf{update} stage can be divided into two separate steps: (i) updating the AoA based on a Discrete Main-Lobe Tracking (ML$_{D}$), and (ii) updating the slope variation.
\begin{figure}
    \centering
    \includegraphics[width=3.15in]{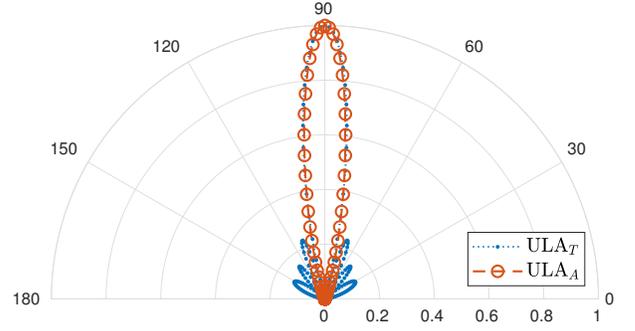}
    \caption{Beam pattern approximation for 8-element ULA}\label{fig:pattern}
\end{figure}
\subsubsection{Discrete Main-Lobe Tracking Algorithm}\label{DMLT}
In this section, we propose a Discrete Main-Lobe Tracking Algorithm, which is consistent with the definition of beam coherence time from \cite{BeamCoh}. This model is built over the assumption that pilot symbol arrives each $T$ within the main-lobe (i.e, within the beam coherence time). This approach is based on the following approximation of the received signal part for ULAs: 
\begin{equation}\label{eqn:approx}
    \left|h(\phi^{}_k,\hat{\phi}^{}_{k-1})\right| \approx \displaystyle e^{-\frac{N^2}{2}[\cos(\phi^{}_k)-\cos(\hat{\phi}^{}_{k-1})]^2}
\end{equation}

A comparison between the true value of the amplitude of the received signal, and the approximation given in \eqref{eqn:approx} can be shown in Fig. \ref{fig:pattern}. We can notice from the figure that the main beam-lobe of the true value $\text{ULA}_T$ almost the same as the approximation $\text{ULA}_A$, and the side-lobes can be neglected especially for a large array size.
The basic idea of the proposed algorithm is that; the value of the signal part $h(\phi^{}_k,\hat{\phi}^{}_{k-1})$: $\mathbb{R}^2\to \mathbb{C}$ is a function of the true value $\phi^{}_k$, and the previous estimate $\hat{\phi}^{}_{k-1}$. If we have an estimate of the signal part from a given measurement, and using the approximation formula in \cref{eqn:approx} we can solve it to find the current estimate. We assume the signal and noise parts are zero-mean Gaussian with variances, $\sigma^2_h$, and $R$ respectively. Also, we assume the signal and noise parts are orthogonal i.e. $\mathbb{E}[h\,n^{}_k]=0$. Hence, we can use the Linear Minimum Mean Estimator (LMMSE) to find $h(\phi^{}_k,{\hat{\phi}^{}_{k-1}})$. Note that LMMSE is one of the most common algorithms used for channel estimation in OFDM based systems \cite{robustnessOfMIMOLMMSE, savaux2017lmmse}. The LMMSE estimate of $h(\phi^{}_k,{\hat{\phi}^{}_{k-1}})$ is found as follows:
\begin{equation}\label{eqn:hest}
\begin{aligned}
    \hat{h}(\phi^{}_k,{\hat{\phi}^{}_{k-1}}) &= y^{}_k\,\displaystyle\frac{\sigma^2_h}{\sigma^2_h+R}\\
    &= y^{}_k\,\displaystyle\frac{\rho}{\rho+1}\\ 
\end{aligned}
\end{equation}
The current estimate is found by solving the approximation given in \cref{eqn:approx}, and the estimate given in \cref{eqn:hest}. An overview of the proposed algorithm is presented in Algorithm \ref{alg1}, which can be described as follows: first, we equalize the estimate from \eqref{eqn:hest} with the approximation formula in \eqref{eqn:approx}, then take $\log$ for both sides. Also, we are going to change the symbol $\phi^{}_k$ in \cref{eqn:approx,eqn:hest} with $\hat{\phi}^{}_k$ since we looking for an estimate.
\begin{equation}\label{eqn:best}
\begin{aligned}
    \ln{|\hat{h}(\hat{\phi}^{}_k,{\hat{\phi}^{}_{k-1}})|}&= -\frac{N^2}{2}[\cos(\hat{\phi}^{}_k)-\cos(\hat{\phi}^{}_{k-1})]^2\\
\end{aligned}
\end{equation}
Hence, the estimate found on \cref{alg:est} is found directly from \cref{eqn:best}. As we notice from Fig. \ref{fig:pattern}, we can have the same amplitude for two different values of AoA, that is why we need to compare the phases of the two possible solutions as found in Algorithm \ref{alg1}. Note that the phase on \cref{alg:phase} is found using $\hat{h}(\phi^{}_k,\hat{\phi}^{}_{k-1})$ from \cref{eqn:hest}.
\begin{algorithm}[t]\small
\setstretch{1.06}
\caption{Discrete Beam Tracking Algorithm}
\label{alg1}
\begin{algorithmic}[1]
% \vspace{-3pt}
\Require $k\ge 1$, $\hat{\phi}^{}_0 = \phi^{}_0$, $T$, and $T^{}_{LR}$
\Comment{{\tiny $T:$Pilot period, $T^{}_{LR}$ link reestablishment time.}}
\Ensure $\hat{\phi}^{}_k$
\While{$kT\le T^{}_{LR}$}
\vspace{2pt}\State $\varphi = \angle{\hat{h}(\phi^{}_k,\hat{\phi}^{}_{k-1})}$, and $\hat{\Omega}^{}_{k-1} = \cos(\hat{\phi}^{}_{k-1})$ \label{alg:phase}
\State {\footnotesize$\hat{\phi}^{}_{k\pm}{=}\arccos{\bigg(\bigg[\hat{\Omega}^{}_{k-1} \pm \sqrt{-\frac{2}{N^2}\ln{\left|\hat{h}(\phi^{}_k,\hat{\phi}^{}_{k-1})\right|}}\bigg]^1_{-1}\bigg)}$}\label{alg:est}
\State $\varphi^{}_{\pm}=\angle{h(\hat{\phi}^{}_{k\pm},\hat{\phi}^{}_{k-1})}$
\If{$\varphi=\varphi^{}_{+}$}
\State $\hat{\phi}^{}_k = \hat{\phi}^{}_{k+}$
\Else
\State $\hat{\phi}^{}_k = \hat{\phi}^{}_{k-}$
\EndIf
\State $k \leftarrow k+1$
\EndWhile
\end{algorithmic}
\end{algorithm}
\subsubsection{Slope Variation Update}
The current estimate of the slope variation ($\hat{\dot{\phi}}^{}_{k}$) is found as an update of $\hat{\dot{\phi}}^{}_{k-1}$ as follows:
\begin{equation}\label{eqn:dotest}
    \hat{\dot{\phi}}^{}_{k} = \hat{\dot{\phi}}^{}_{k-1} + \displaystyle\frac{Q\,T\,\xi\,\Im{[y^{}_k]}}{Q\,T\,\xi^2 + 0.5\,R},
\end{equation}
\noindent where $\xi = \pi\,T\,\sin(\hat{\phi}^{}_{k-1})\,\frac{N-1}{2}$, and we assume initial estimate $\hat{\dot{\phi}}^{}_0 = 0$. As shown from \cref{eqn:dotest}, the current estimate depends on the previous estimate and an update term. The update term is found as follows; first, we use a first order approximation of the signal part in the measurement equation by:
\begin{equation}
    h(\dot{\phi}^{}_k,\hat{\phi}^{}_{k-1})= h(\hat{\dot{\phi}}^{}_{k-1},\hat{\phi}^{}_{k-1}) + \varepsilon^{}_k\,h^{'}(\hat{\dot{\phi}}^{}_{k-1},\hat{\phi}^{}_{k-1}),
\end{equation}
\noindent where $\varepsilon^{}_k = (\dot{\phi}^{}_k - \hat{\dot{\phi}}^{}_{k-1})$, and by assuming good estimate (i.e., $\hat{\dot{\phi}}^{}_{k-1}$,$\hat{\phi}^{}_{k-1}$) we can have $h^{'}(\hat{\dot{\phi}}^{}_{k-1},\hat{\phi}^{}_{k-1}) \approx j\varepsilon^{}_k\,\xi$, and $h(\hat{\dot{\phi}}^{}_{k-1},\hat{\phi}^{}_{k-1})\approx 1$. In that case, the imaginary part of the measurement equation is given by:
\begin{equation}
    \Im{[y^{}_k]} = \varepsilon^{}_k\,\xi + \Im{[n^{}_k]}/\rho,
\end{equation}
\noindent where $\varepsilon^{}_k \sim \mathcal{N}(0,Q\,T)$, and for a given $\xi$, the MMSE of $\varepsilon^{}_k$ (i.e., $\hat{\varepsilon}^{}_k$) is given by:
\begin{equation}
\begin{aligned}
\hat{\varepsilon}^{}_k &= \mathbb{E}\left[\varepsilon^{}_k|y^{}_k\right]\\
&=\displaystyle\frac{Q\,T\,\xi\,\Im{[y^{}_k]}}{Q\,T\,\xi^2 + 0.5\,R}\\
\end{aligned}    
\end{equation}

\section{Design Parameters}\label{sec:LRT}
The main goal of beam tracking, is to elongate the link reestablishment time $T^{}_{LR}$; the time at which we lost the tracking of the AoA, and we need to search for the new best incoming path. Recall the frame structure in Fig. \ref{fig:Frame}, where we need to send a frame of $k$ time slots with a fixed rate. 
In Section \ref{sec:Nres1}, we show that large array size is not able to cope with channel variation for a given \textit{pilot period} (Time slot) $T$. Hence, we need to choose the \textit{pilot period} $T$ based on array size beside other factors we will discuss later. In the following, we propose two different ways to choose the \textit{pilot period}; (i) comparable to the beam coherence and (ii) to maintain specific outage probability and fixed rate.

\subsection{Beam Coherence Time}
Intuitively, the \textit{pilot period} $T$ has to be comparable to the beam coherence time $T^{}_b$ to cope with the channel variation. From \cite{BeamCoh}, the beam coherence time is defined as the time at which the power received at perfect alignment time drop by $\zeta$ at time $t+T^{}_b$.
\begin{equation}\label{eqn:zeta}
    \displaystyle\frac{P(t+T^{}_b)}{P(t)} = \zeta
\end{equation}
Since we assume perfect alignment at time $t$, then the received power $P(t) = 1$. For ULA, the received power at time $t+T^{}_b$ while ignoring the noise power is given by:
\begin{equation}\label{eqn:PR}
\begin{aligned}
    P(t+T^{}_b) &= \left|\boldsymbol{w}^{\dag}(\phi(t))\boldsymbol{a}^{}_R(\phi(t+T)b))\right|^2\\
    &{=}\displaystyle\frac{1}{N^2}\left|\sum_{m=0}^{N-1}e^{-j2\pi m\Delta(\cos(\phi(t+T^{}_b))-\cos(\phi(t)))}\right|^2\\
\end{aligned}
\end{equation}

\begin{figure}[t]
 \centering
\includegraphics[scale=0.7]{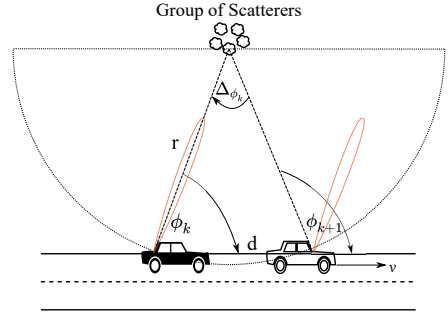}
 \caption{Model of variation over pilot period}\label{fig:motion}
\end{figure}

Now, we try to visualize the picture of a simple example to grasp the whole idea. As shown in Fig. \ref{fig:motion}, we assume the receiver is perfectly aligned with the incoming path at the first position. The receiver is moving with speed $v$ and a fixed beam direction equals to the AoA at the first position, i.e. $\phi^{}_k$. Hence, after motion for a distance $d$ there will be a $\Delta_{\phi^{}_k}$ misalignment, and the received power will be dropped to $\zeta$. Our model relates the channel variation to the angular variation $\dot{\phi}^{}_k$, and Fig. \ref{fig:motion} shows how we relate the linear speed $v$ to that angular variation as:
\begin{equation}
    \begin{aligned}
        \phi^{}_{k+1}&=\phi^{}_k + T\,\dot{\phi}^{}_k\\
        &=\phi^{}_k + T\,\frac{v\,\sin(\phi^{}_k)}{r},\\
    \end{aligned}
\end{equation}
\noindent where $r$ is the distance from the source of the path to the receiver and now we can say that the angular variation is bounded by speed over the distance as : $\dot{\phi}^{}_k \le \frac{v}{r}$.
This is the typical situation for the Discrete tracking approach, and we are seeking the \textit{pilot period} $T$ to be within the beam coherence time $T^{}_b$. For a Continuous-Discrete tracking algorithm, the beam direction changes dynamically based on estimate of the slope variation $\dot{\phi}^{}_k$, then to find the pilot duration we need to take care of slope variation estimate in addition to the AoA variations. In this case, the notion of beam coherence time does not hold for the Continuous-Discrete algorithms, and we introduce the notion of the \textit{Beam Locking Time} $T^{}_L$. This time represents the duration at which the power will drop by $\zeta$ while the receiver continuously update the estimate of the AoA and the beamforming. 

In order to find a consistent formula of the \textit{pilot period}, we assume it is the time to have an average drop $\mu^{}_{\zeta}$ {in the received power} instead of $\zeta$ since both sides of \cref{eqn:zeta} are random. {Furthermore, we incorporate a design parameter $\kappa$ to contemplate any estimation errors in the slope variations. This design parameter, $0 < \kappa < 1$ reflects the confidence level in the predicted slope variation, where a smaller value indicates higher confidence. For instance, for $\kappa = 0$, the system designer indicates full confidence in the predicted slope value, and no extra measurement will be required. On the other hand, $\kappa=1$, indicates a total lack of confidence in the predicted slope value and a new measurement will be required every beam coherence time ($T^{}_{b}$)}. 
Now, we introduce how to choose the \textit{pilot period} for a Discrete and Continuous-Discrete beam tracking using Theorem \ref{th:TBD}, {and its proof can be found in Appendix \ref{apx:th1}, and \ref{apx:th2}.}
% \vspace{-0.5em}
\begin{theorem}\label{th:TBD}
For the channel model with angular variation given by \cref{eqn:phit}, the link reestablishment interval $T^{}_{LR}$ is achievable if the \textit{pilot period} $T\le T^{}_{L}$ is given by:
\begin{equation}\label{th:TBCD}
    T^{}_L = \sqrt{\displaystyle\frac{1/\mu^2_{\zeta}-1}{2\,\kappa\,Q\,T^{}_{LR}\,\sin^2(\phi^{}_k)N^2}}.
\end{equation}
\end{theorem}
Clearly, the power will drop in the discrete beam tracking algorithm faster than the continuous-discrete tracking algorithms. This is because continuous-discrete tracking continuously varies the beamforming based on the prediction of the AoA variation and gain from discrete pilot updates. While the discrete approaches only update the AoA upon the arrival of pilot symbols.
\subsection{Outage Probability}
For the situation where the transmitter is aware of the channel distribution, this motivates the transmitter to choose the \textit{pilot period} that sustain a specific outage probability $P^{}_{out}$ for a fixed rate $\mathcal{R}^{}_{f}$. In that case, the \textit{pilot period} is chosen based on Theorem \ref{th:TDout}.
% \vspace{-0.5em}
\begin{theorem}\label{th:TDout}
For the channel model with angular variation given be \cref{eqn:phit}, the link reestablishment interval $T^{}_{LR}$ is achievable with outage probability $P^{}_{out}$, and a fixed rate $\mathcal{R}^{}_{f}$ if the \textit{pilot period} $T\le T^{}_o$ is given by:
\begin{equation}\label{eqn:Poutcd}
T^{}_{o} = \sqrt{\displaystyle\frac{\log(\rho/(2^{\mathcal{R}^{}_{f}}-1))}{N^2\sin^2(\phi^{}_k)\kappa\,QT^{}_{LR}(\mathbb{Q}^{-1}(P^{}_{out}/2))^2}}.
\end{equation}
\end{theorem}
Similarly, $0<\kappa<1$, and for $\kappa =1$, then $T^{}_{o}$ is valid as the \textit{pilot period} for the Discrete 
% beam tracking algorithm
tracking framework, otherwise, $T^{}_{o}$ is valid for the Continuous-Discrete beam tracking algorithm.
% \vspace{-0.5em}
\begin{proof}\label{proof:thout}
From Appendix \ref{apx:th3}, we get the outage probability for a Continuous-Discrete tracking approach is given by \cref{eqn:CDFCD}:
\begin{equation}\label{eqn:pout}
    \begin{aligned}
    P^{}_{out}&=P(\gamma\le\gamma^{}_k)\\    &=2\,\mathbb{Q}\left(\sqrt{\displaystyle\frac{\log(\rho/\gamma^{}_k)}{N^2{T^{}_o}^2\sin^2(\phi^{}_k)\kappa\,QT^{}_{LR}}}\right)\\
    \end{aligned}
\end{equation}
\noindent for a given $\gamma^{}_k$, the corresponding spectral efficiency is $\mathcal{R}^{}_{f}=\log^{}_2(1+\gamma^{}_k)$. Alternatively, we can replace $\gamma^{}_k$ in \cref{eqn:pout} by $2^{\mathcal{R}^{}_{f}}-1$. After that, we divide both sides by $2$, then take the inverse of the Q function. Finally, \cref{eqn:Poutcd} is found directly by taking $T^{}_o$ to the left side.
Similarly, The Discrete case where $\kappa=1$ is derived in similar procedures as \cref{eqn:Poutcd} by starting from the CDF for the Discrete case which is given by \cref{eqn:CDFD}.
% \vspace{-11pt}
\end{proof}
\subsection{Pilot Overhead Reduction} The key point of utilizing the spatial variations in-between two pilot symbols is to reduce pilot overhead by extending the pilot symbols duration. Now, we show how a Continuous-Discrete tracking algorithms reduce pilot overheads as follows:
\subsubsection{Beam Coherence and Locking Time}
% \vspace{-10pt}
\begin{equation}
    \begin{aligned}
        \frac{T^{}_b}{T^{}_L} & = \sqrt{\displaystyle\frac{1/\mu^2_{\zeta}-1}{2\,Q\,T^{}_{LR}\,\sin^2(\phi^{}_k)N^2}\displaystyle\frac{2\,\kappa\,Q\,T^{}_{LR}\,\sin^2(\phi^{}_k)N^2}{1/\mu^2_{\zeta}-1}}\\
        &=\sqrt{\kappa}
    \end{aligned}
    % \vspace{-10pt}
\end{equation}
% \begin{figure*}
% \img{Lup/PHI_SNR_20_Q_10000_T_0.0025_HS_ee.eps}{AoA tracking with 16, 64 ULAs for $T=2.5$ ms.}{fig:phi}
% % {Angle of Arrival Tracking with 16 and 64 ULAs for $T=2.5\,ms$.}{fig:phi}
% \img{Lup/MSE_SNR_20_Q_10000_T_0.001_t.eps}{MSE for different approaches using 16, and 64 ULAs}{fig:MSEant}
% % {MSE for different tracking approaches using 16, and 64 ULAs}{fig:MSEant}
% \img{Lup/MSE_SNR_20_10_Q_10000_T_0.001_N_64_t.eps}{MSE for different approaches for 20, and 10 dB}{fig:MSEdB}
% % {MSE for Different Tracking Approaches for 20, and 10 dB}{fig:MSEdB}
% % \vspace{-1.5em}
% \end{figure*}
\subsubsection{Outage Probability Time}
% \vspace{-10pt}
\begin{equation}
    \begin{aligned}
        \frac{T^{}_{o}{\text{\tiny$(D)$}}}{T^{}_{o}{\text{\tiny$(CD)$}}}&= {\text{\scriptsize$\sqrt{\displaystyle\frac{\log(\frac{\rho}{(2^{\mathcal{R}^{}_{f}}-1)})\, N^2\sin^2(\phi^{}_k)\kappa\,QT^{}_{\text{\tiny$LR$}}(\mathbb{Q}^{-1}(\frac{P^{}_{out}}{2}))^2}{N^2\sin^2(\phi^{}_k)QT^{}_{{\text{\tiny$LR$}}}(\mathbb{Q}^{-1}(\frac{P^{}_{out}}{2}))^2\,\log(\frac{\rho}{(2^{\mathcal{R}^{}_{f}}-1)})}}$}}\\
        &=\sqrt{\kappa}
    \end{aligned}
\end{equation}

Hence, utilizing the Continuous-Discrete algorithms results in overhead reductions by $(1-\sqrt{\kappa})\%$.
\subsection{Effective Achievable Rate}
Now, we are looking to represent the overhead reduction by the effective achievable rate by excluding the time of pilots training. The effective rate is given by:
\begin{equation}\label{eqn:reff}
    \begin{aligned}
        \mathcal{R}^{}_{e} &= \eta\,\mathcal{R}^{}_{a}\\
        &=\frac{T-T^{}_s}{T}\times\,\frac{T^{}_{LR}}{T^{}_{LR}+T^{}_{sw}}\,\mathcal{R}^{}_{a},\\
    \end{aligned}
\end{equation}
\noindent where $\mathcal{R}^{}_{a}$ is the achievable rate, $T$ will be replaced by $T^{}_b$, $T^{}_L$, $T^{}_{o}$(D), and $T^{}_{o}$(CD) for each case of choosing the \textit{pilot period}. Also, $T^{}_s$ is the training time for a single pilot symbol, and $T^{}_{sw}$ is the beam sweeping time, which is needed to find the angle of arrival after losing the tracking. Considering the sweeping time, we utilize the same formula proposed by \cite{BeamCoh} based on the hierarchical beam code book in \cite{beamsweep} and follows the IEEE 802.15.3c guidance. The formula was given as function of the beamwidth, which can be shown as function of the array size as:
\begin{equation}
    T^{}_{sw} = \mathbb{L}\,(\pi\,N)^{(2/\mathbb{L})}\,T^{}_s,
\end{equation}
\noindent where $\mathbb{L}$ is the number of levels in the hierarchical beam codebook given in \cite{beamsweep}.
\section{Numerical Results}\label{sec:Nres}
\subsection{Performance of Beam Tracking Algorithms}\label{sec:Nres1}
In this section, we evaluate and compare the performance of EKF and FBT algorithms under both the Continuous-Discrete and the Discrete frameworks. Recall that algorithms under our proposed continuous-discrete framework are given the subscript ``CD'', while state-of-the-art discrete algorithms are given the subscript ``D''. Our proposed ML algorithm with both flavors ML$_{CD}$ and ML$_{D}$ will be studied in Section \ref{sec:Nres2}. For each experiment, we assume perfect knowledge of the initial AoA at $t=0$, i.e., $\phi^{}_0$.
We simulate $\dot{\phi}(t)$ as a Brownian motion and $\phi(t)$ as in \eqref{eqn:phit}, and we let the algorithms run for a total tracking time of 100 ms and average the output over 5000 runs.
Our performance metrics are: (1) the MSE of the estimated AoA, and (2) the average received SNR. Except when stated otherwise, we let $Q=10^4$, $N=64$ ULA and we fix $\rho=20$ dB. We evaluate the performance under two array orientations, i.e., ULA and UPA.
\begin{figure}
    \centering
    \includegraphics[scale=0.6]{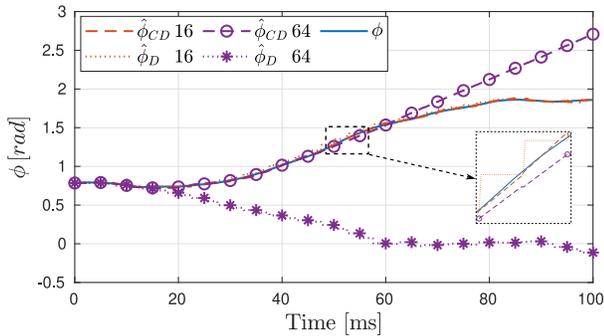}
    \caption{AoA tracking with 16, 64 ULAs for $T=2.5$ ms.}
    \label{fig:phi}
\end{figure}
\begin{figure}
    \centering
    \includegraphics[scale=0.6]{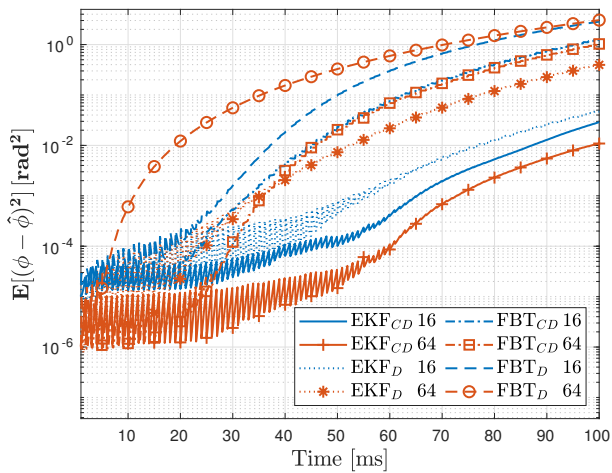}
    \caption{MSE for different approaches using 16, and 64 ULAs}
    \label{fig:MSEant}
\end{figure}

\noindent\textbf{Different ULA Array Size:} Fig. \ref{fig:phi} depicts a sample path of the progression of the simulated true value of the AoA as well as its tracking performance using both EKF$_{CD}$ and EKF$_D$ algorithms when ULAs of sizes 16 and 64 are employed. The \textit{pilot period} is fixed at $T = 2.5$ ms. For the ULA of size 16, both EKF$_{CD}$ and EKF$_{D}$ frameworks provide accurate tracking performance for the entire tracking duration. Only a closer inspection reveals the superiority of EKF$_{CD}$ as shown in the zoomed-in part. However, as the array size gets larger, the beamwidth gets narrower, making it harder to track the beam. Consequently, for ULAs with 64 antenna elements, we see a clear difference in favor of EKF$_{CD}$ where it is able to track the beam for a significantly longer time duration ($\sim$200\% longer), compared to EKF with discrete updates. In general, as the number of antenna elements increases, the performance gap between EKF$_{CD}$ and EKF$_D$ increases in favor of our EKD$_{CD}$ solution as long as the variation in the AoA over $T$ is less than the variation over the beam coherence time.

In Fig. \ref{fig:MSEant}, the MSE for the continuous-discrete and the discrete approaches for $T=1.0$ ms is shown. Here, we also provide results for the FBT$_{CD}$ and FBT$_D$ algorithms. These results show that the continuous-discrete framework significantly improves performance over the baseline discrete solutions. We also see that EKF$_{CD}$ has the lowest MSE, i.e., best tracking performance, among all other solutions.
Interestingly, EKF$_{CD}$ performs even better for the larger array size, despite the overall difficulty in tracking sharper beams.
On the other hand, the performance of EKF$_{D}$ degrades by increasing the array size (especially when AoA variations become faster over the tracking time).
This happens since the larger array size has a shorter beam coherence time, and EKF$_{D}$ becomes unable to cope with the fast channel variation.
The Effect of the array size over FBT$_{CD}$ and FBT$_{D}$ is similar to that of EKF$_{CD}$ and EKF$_{D}$. Yet, EKF$_{D}$ and EKF$_{CD}$ outperform the Fast-tracking approach (especially when the channel variations become faster).
EKF$_{CD}$ achieves $80\%$ and $99.5\%$ lower MSE than EKF$_{D}$, for 16 and 64 array sizes, respectively.
\begin{figure}[t]
    \centering
    \includegraphics[scale=0.6]{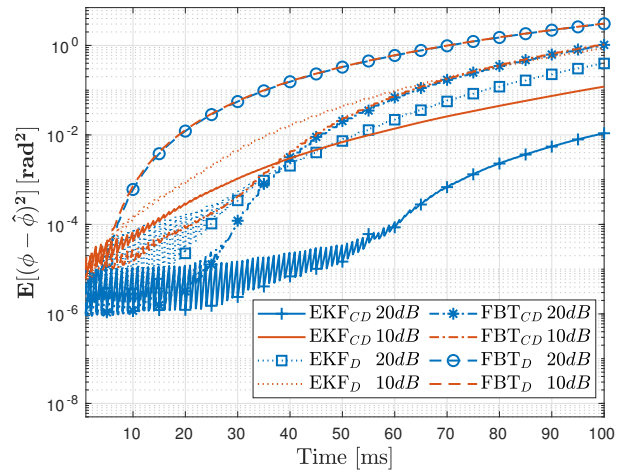}
    \caption{MSE for different approaches for SNR = 20, 10 dB.}
    \label{fig:MSEdB}
\end{figure}
% \begin{figure*}
% \img{Lup/MSE_SNR_20_Q_1000_T_0.001_0.0025_N_64_t.eps}{MSE for different $T=1.0,\,2.5$ ms}{fig:MSE1000}
% % {MSE for Different Tracking Approaches for $T=1.0,\,2.5\,ms$}{fig:MSE1000}
% \img{Lup/ASNR_20_Q_10000_T_0.001_0.0025_t.eps}{SNR for ULA for different array sizes}{fig:SNR16}
% % {Average SNR for ULA for Different Array Sizes}{fig:SNR16}
% \img{Lup/ASNR_20_Q_10000_T_0.001_0.0025_G_t.eps}{SNR for ULA including array gain}{fig:SNR2_5}
% % \vspace{0.5em}
% % {Average SNR for ULA for Different Array Sizes}{fig:SNR2_5}\vspace{0.5em}
% \img{Lup/ASNR_20_Q_10000_T_0.001_0.0025_P_t.eps}{SNR for UPA, ULA for different array sizes}{fig:SNRPNG}
% % {Average SNR for UPA, ULA for Different Sizes}{fig:SNRPNG}
% % {Average SNR for UPA, ULA for Different Array Sizes}{fig:SNRPNG}
% \img{Lup/ASNR_20_Q_10000_T_0.001_0.0025_PG_t.eps}{SNR for UPA, ULA including array gain}{fig:SNRPG}
% % {Average SNR for UPA, ULA for Different Sizes}{fig:SNRPG}
% % {Average SNR for UPA, ULA for Different Array Sizes}{fig:SNRPG}
% \img{Lup/SNR_10dB_4_256_N_bits_MN.eps}{Average SNR for different quantization bits}{fig:SNRQuan}
% % \vspace{-1.5em}
% \end{figure*}

\noindent\textbf{Different SNR values:} Fig. \ref{fig:MSEdB} shows the MSE under different SNR values; 10 and 20 dB. Here, we fix the array size to 64 and the {pilot} period to $T=1.0$ ms. As expected, decreasing the SNR degrades the performance for all approaches. EKF$_{CD}$, however, still outperforms all other algorithms due to its ability to predict the rate of change of the AoA between pilot symbols. The superiority of EKF$_{CD}$ is very dominant that even at 10 dB, it still has comparable performance to FBT$_{CD}$ and EKF$_{D}$ with 20 dB. The MSE for all approaches is tiny at the beginning of the tracking period since the variation of AoA is negligible. As the tracking time advances, the variation in AoA increases, and the MSE for all approaches increases.
\begin{figure}
    \centering
    \includegraphics[scale=0.6]{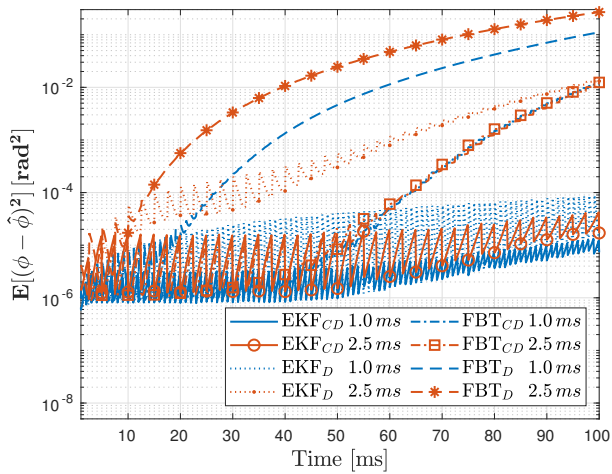}
    \caption{MSE for different pilot periods: $T=1.0$, and $2.5$ ms}
    \label{fig:MSE1000}
\end{figure}
\begin{figure}
    \centering
    \includegraphics[scale=0.6]{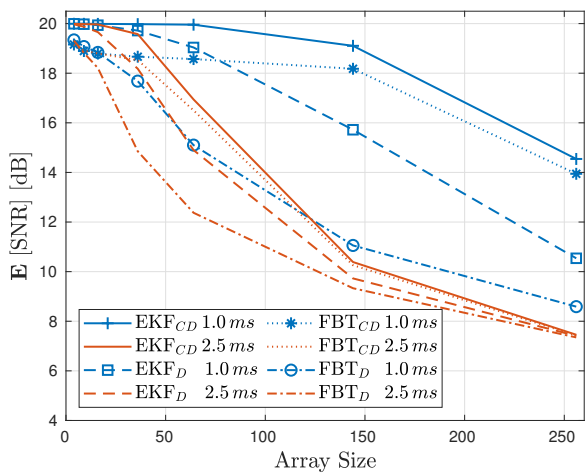}
    \caption{SNR for ULA for different array sizes}
    \label{fig:SNR16}
\end{figure}

\noindent\textbf{Different Pilot Frequencies:} Measurement frequency is an important design parameter. More frequent pilots ensure better beam tracking, but it also wastes more transmission opportunities. Here, we shed more light on the effect of tracking performance as the measurement frequency changes, where we plot the MSE for a medium AoA variation at $Q=10^3$ for a 64-element ULA, as shown in Fig. \ref{fig:MSE1000}.
First, we observe that for $T = 1.0$ ms, the performance of both EKF$_{CD}$ and EKF$_D$ are almost identical at the measurement update instances. However, the MSE of EKF$_D$ fluctuates more aggressively in-between measurements. This is due to the AoA estimate of the EKF$_{D}$ being kept constant between measurements despite the continuous AoA changes over time. Second, accounting for the nature of variation between measurements gives an advantage to EKF$_{CD}$ and allows it to use less frequent measurements while keeping the MSE comparable to EKF$_{D}$. This can be seen when comparing the MSE of EKF$_{CD}$ with $T=2.5$ ms to the MSE of EKF$_{D}$ with $T=1.0$ ms. Finally, at the same $T$, FBT$_{CD}$ achieves up to $99.8\%$ lower MSE compared to FBT$_{D}$.
Also, FBT$_{CD}$ performs almost identically at both $T=1.0$ and $2.5$ ms, showing the effectiveness of continuous AoA updates.

\begin{figure}
    \centering
    \includegraphics[scale=0.6]{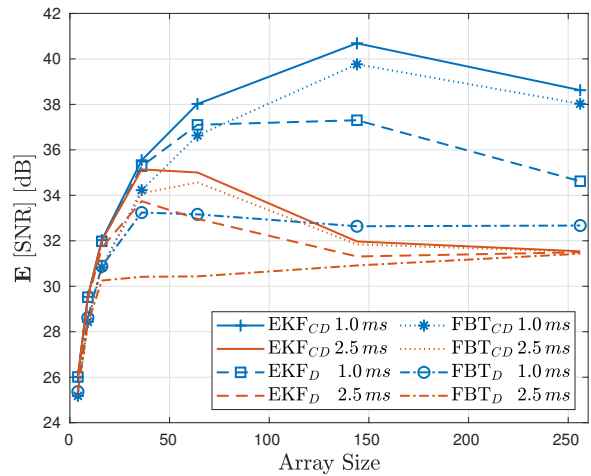}
    \caption{SNR for ULA including array gain}
    \label{fig:SNR2_5}
\end{figure}

In Fig. \ref{fig:SNR16}, we present the average received SNR for different $T$.
Averaging is over the whole tracking time and 5000 runs.
Here, we also normalize $\rho$, with respect to the array size. This normalization helps isolate the tracking performance as a function of the beamwidth only.
Next in Fig. \ref{fig:SNR2_5}, we will remove this normalization and study the overall performance with sharp beams, by accounting for the beam gain, as well.
We notice that average SNR is degraded by increasing the array size since increasing the array size decreases the beam coherence time. In order to have good performance over large array sizes, we need to utilize smaller $T$. The EKF$_{CD}$ outperforms the EKF$_{D}$ by 1 dB and up to 4 dB for small and large array sizes, respectively, while the FBT$_{CD}$ has an advantage by 4 dB and can reach 7 dB. Increasing the \textit{pilot period} degrades the SNR for all approaches, yet EKF$_{CD}$ still outperforms EKF$_{D}$ by 2 dB for $T=2.5$ ms. Also, FBT$_{CD}$ has superiority over FBT$_{D}$ by 5 dB for $T=2.5$ ms.

Fig. \ref{fig:SNR2_5} shows the average SNR for 16 dB per each antenna element. We observe that increasing the array size increases the average received SNR to a certain point, before it drops slowly for larger sizes. This can be explained by arguing that the array gain compensates for small mismatches in beam alignment, despite the difficulty imposed by the larger array size. However, as arrays get larger, beams get sharper, and alignment fails more often, leading to a drop in the SNR.
\begin{figure}
    \centering
    \includegraphics[scale=0.6]{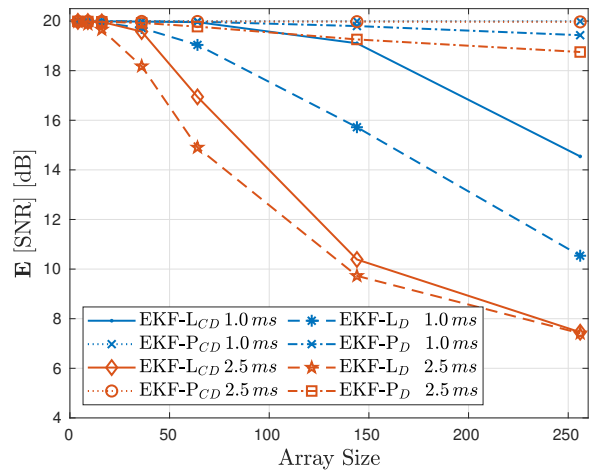}
    \caption{SNR for UPA, ULA for different array sizes}
    \label{fig:SNRPNG}
\end{figure}

\noindent\textbf{Uniform Planar Arrays:}
A UPA has better spatial properties than ULA in terms of beam coherence time. Hence, a UPA can utilize longer $T$ and still obtain better performance than ULA. We notice from Fig. \ref{fig:SNRPNG} that UPA scenarios, which are denoted by EKF-P$_{CD}$ and FBT-P$_{CD}$ have better average SNR for $T=1.0,\,2.5$ ms, especially for large array sizes and fast channel variations.
\begin{figure}
    \centering
    \includegraphics[scale=0.6]{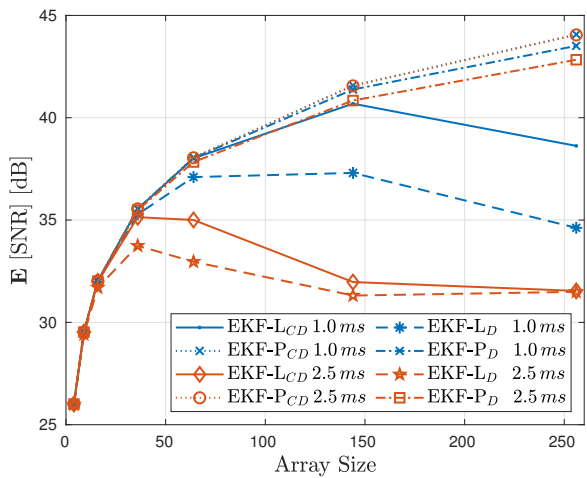}
    \caption{SNR for UPA, ULA including array gain}
    \label{fig:SNRPG}
\end{figure}
\begin{figure}
    \centering
    \includegraphics[scale=0.6]{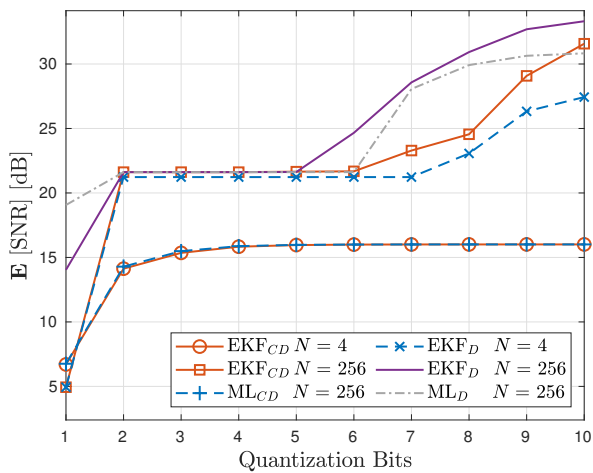}
    \caption{Average SNR for different quantization bits}
    \label{fig:SNRQuan}
\end{figure}

Fig. \ref{fig:SNRPG} shows the effect of the array gain, similar to Fig. \ref{fig:SNR2_5}. The UPA shows a better use of the array gain, even for the large array size. Intuitively, the average SNR for UPA will eventually start dropping when the beam coherence time of larger arrays becomes much smaller than $T = 1.0,\,2.5$ ms.

{\noindent\bf Phase-shifters Resolution:} \cref{fig:SNRQuan} shows the effect of the phase-shifters resolution. We show the average SNR under two different array sizes. It is clear that for a small array size (i.e., $N=4$), the effect of resolution saturates for a small number of bits. Also, there is no gain from the continuous-discrete approach since the beam is too wide, which simplifies the tracking problem and diminishes the resolution effect. Increasing the array size gives an advantage to discrete approaches for the very low resolution since it is only affected by resolution errors upon pilot arrival. However, increasing the resolution makes the continuous-discrete approaches outperform the discrete solutions due to increasing the possible directions that can be utilized.
\begin{figure}
    \centering
    \includegraphics[scale=0.23]{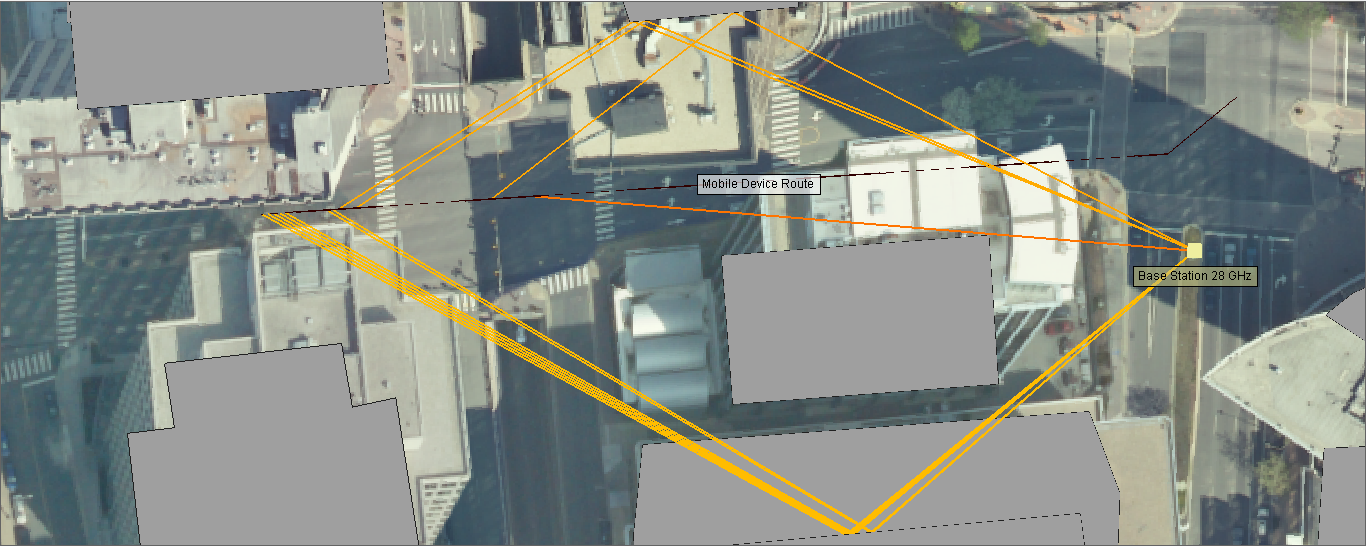}
    \caption{Sample route of the UE in urban city}
    \label{fig:WIroute}
\end{figure}
\begin{figure}
    \centering
    \includegraphics[scale=0.6]{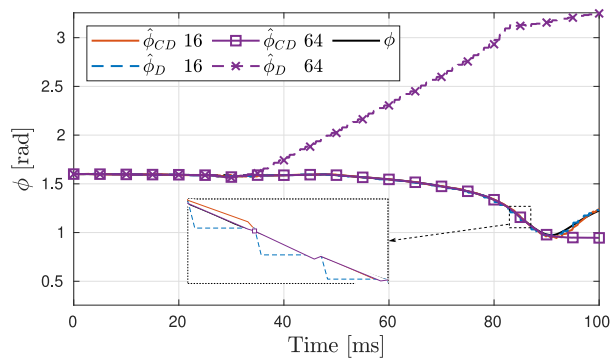}
    \caption{AoA tracking for a practical sample path}
    \label{fig:WIAoA}
\end{figure}

\begin{figure}
    \centering
    \includegraphics[scale=0.6]{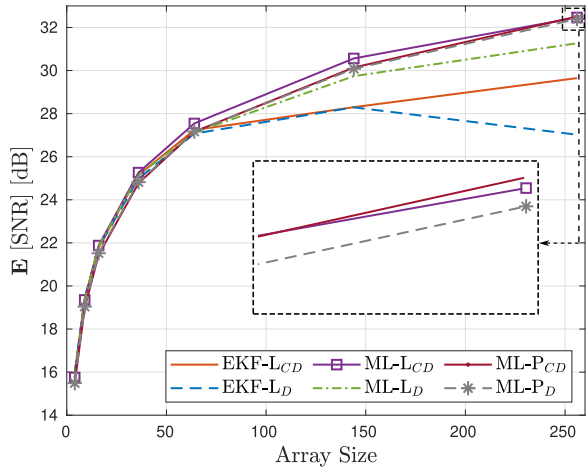}
    \caption{Average SNR for tracking in a practical situation}
    \label{fig:WISNR}
\end{figure}

{\noindent\bf Wireless Insite:} Here, we provide simulation results with a tool that simulates the propagation of radio waves in real situations. We choose the urban area (Rosslyn City) and assume a person is moving over different paths between two points as shown in Fig. \ref{fig:WIroute}. Both the BS and UE operate at 28 GHz, and the UE sometimes is in LOS or Non-LOS or blockage situations at any point over the route. Beam tracking considers the mild misalignment in beam direction over the frame length ($T_{LR}^{}$), any abrupt changes due to blockage or other reasons should be solved through beam realignment. Fig. \ref{fig:WIAoA} depicts that the continuous-discrete approach still outperforms the discrete solutions for a practical AoA variation similar to the discussion in Fig. \ref{fig:phi}. Finally, Fig. \ref{fig:WISNR} presents the effect of considering the variation in both the azimuth and elevation planes. The tracking problem in the elevation plane is much easier for the ULA since it has equal gain in the elevation plane. The ML-L$_{CD}$ (ULA) approach outperforms the ML-P$_{CD}$ (UPA) since the estimation errors in both planes affect the UPA performance. Nevertheless, increasing the array size changes the situation in favor of the UPA, as shown in the zoomed-in part.
\begin{table*}
\centering
\caption{\scriptsize\textsc{pilot overhead reduction under two different methods of choosing $T$ for different antenna array size with $T^{}_{LR}=100$} ms}
\label{tab:overhead}
\centering
\resizebox{0.9\textwidth}{!}{%
\begin{tabular}{|c|c|c|c|c|c|c|c|c|c|c|c|} 
\cline{2-12}
\multicolumn{1}{c|}{} & \multicolumn{11}{c|}{Overhead Reduction} \\ 
\hhline{~-----------|}
\multicolumn{1}{c|}{} & \multicolumn{2}{c|}{{\cellcolor[rgb]{0.875,0.875,0.875}}Array Size } & {\cellcolor[rgb]{0.875,0.875,0.875}}4 & {\cellcolor[rgb]{0.875,0.875,0.875}}9 & {\cellcolor[rgb]{0.875,0.875,0.875}}16 & {\cellcolor[rgb]{0.875,0.875,0.875}}36 & {\cellcolor[rgb]{0.875,0.875,0.875}}64 & {\cellcolor[rgb]{0.875,0.875,0.875}}144 & {\cellcolor[rgb]{0.875,0.875,0.875}}256 & {\cellcolor[rgb]{0.875,0.875,0.875}}512 & {\cellcolor[rgb]{0.875,0.875,0.875}}1024 \\ 
\hline
{\cellcolor[rgb]{0.882,0.941,1}} & \multicolumn{2}{c|}{{\cellcolor[rgb]{0.882,0.941,1}}Beam Coherence} & 19\% & 29\% & 36\% & 46\% & 53\% & 62\% & 67\% & 72\% & 76\% \\ 
\hhline{|>{\arrayrulecolor[rgb]{0.882,0.941,1}}->{\arrayrulecolor{black}}-----------|}
{\cellcolor[rgb]{0.882,0.941,1}} & {\cellcolor[rgb]{0.882,0.941,1}} & {\cellcolor[rgb]{0.882,0.941,1}}$0.5\,\mathcal{R}_{max}$ & 19\% & 29\% & 35\% & 44\% & 52\% & 60\% & 65\% & 71\% & 75\% \\ 
\hhline{|>{\arrayrulecolor[rgb]{0.882,0.941,1}}-->{\arrayrulecolor{black}}----------|}
\multirow{-3}{*}{{\cellcolor[rgb]{0.882,0.941,1}}  How to choose T} & \multirow{-2}{*}{{\cellcolor[rgb]{0.882,0.941,1}}Outage Probability} & {\cellcolor[rgb]{0.882,0.941,1}}$0.95\,\mathcal{R}_{max}$ & 60\% & 60\% & 61\% & 73\% & 76\% & 80\% & 83\% & 85\% & 87\% \\
\hline
\end{tabular}%
}
% \vspace{-0.5em}
\end{table*}
\begin{figure}
    \centering
    \includegraphics[scale=0.6]{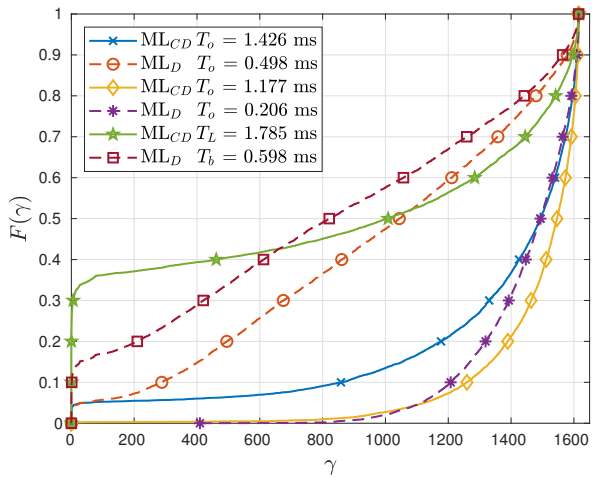}
    \caption{CDF of the received SNR}
    \label{fig:CDFSNR}
\end{figure}

\subsection{Choosing the pilot period}\label{sec:Nres2}
Here, we verify our discussion of choosing the \textit{Pilot Period} $T$ using the proposed tracking algorithms ML$_{CD}$, and ML$_{D}$ since these algorithms are modeled on the concept of beam coherence.
The assumptions that we assumed in Section \ref{sec:Nres1} still hold. In addition, we assume $\phi^{}_0=\pi/2$, $Q=10^3$, the frame length $T^{}_{LR} = 100$ ms, and SNR/antenna element 8 dB. Finally, the average drop in the SNR $\mu^{}_{\zeta} = 0.5$ at time $T^{}_{LR}$, the outage probability $P^{}_{out} = 0.05$, and a fixed rate $\mathcal{R}^{}_{f} = \delta\, \mathcal{R}^{}_{max}$ as a factor of the maximum rate  $\mathcal{R}^{}_{max} = \log^{}_2(1+\rho)$.
Our performance metrics are: (1) the CDF of the received SNR and the achievable rate, and (2) the effective achievable rate.

First, we numerically investigate the pilot overhead reduction, as shown in Table \ref{tab:overhead}. We compare the overhead reduction for different array sizes by choosing the \textit{pilot period} $T$ either by the beam coherence or outage probability definitions. The overhead reduction increases with the array size, and this comes from the fact that for a small array size, the beam is wide, and the ML$_{D}$ can cope with channel variation with a \textit{pilot period} comparable to ML$_{CD}$. Yet, for a large array size, the beam coherence time becomes much smaller, and the \textit{pilot period} becomes smaller for a ML$_{D}$, which gives the advantage to the ML$_{CD}$. Increasing the rate to $0.95\mathcal{R}^{}_{max}$, imposes the tracking algorithms to sustain higher received SNR, which can be maintained by decreasing the \textit{pilot period}. The ML$_{CD}$ algorithm can sustain a higher rate with a very comparable \textit{pilot period} used at the lower rate. On the contrary, the ML$_{D}$ needs to reduce the \textit{pilot period} too much. This can be noticed from Table \ref{tab:overhead}, where the overhead reduction is increased even for the small array sizes. Finally, the overhead reduction for beam coherence and the outage probability with the rate $\mathcal{R} = 0.5\mathcal{R}^{}_{max}$ are almost the same, but both have different \textit{pilot periods}. 
\begin{figure}
    \centering
    \includegraphics[scale=0.6]{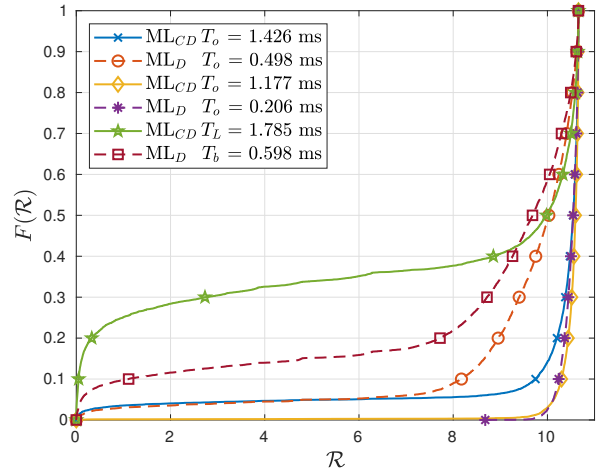}
    \caption{CDF of the achievable rate}
    \label{fig:CDFR}
\end{figure}

Now, we consider the CDF to compare between the two proposed methods of choosing the \textit{pilot period} $T$. In Fig. \ref{fig:CDFSNR}, we present the CDF of the received SNR for array size $N=256$ for three different cases: (1) outage with $\mathcal{R}^{}_{f} = 0.5\,\mathcal{R}^{}_{max}$ (2) outage with $\mathcal{R}^{}_{f} = 0.95\,\mathcal{R}^{}_{max}$ (3) beam coherence with $\mu^{}_{\zeta}=0.5$. For case 1, with \textit{pilot period} $T^{}_o = 1.426$, and 0.498 ms for ML$_{CD}$, and ML$_{D}$ respectively. It is clear ML$_{CD}$ outperforms the ML$_{D}$, where it can maintain SNR up to 600 for the same outage probability. Case 2, with \textit{pilot period} $T^{}_o = 1.177$, and 0.206 ms still shows superiority in terms of SNR but not significant as in case 1. However, the pilot overhead reduction is significantly increased in case 2 to 83\% instead of 65\% as in case 1. Case 3 shows that both ML$_{CD}$ and ML$_{D}$ nearly have the same average SNR as designed but the ML$_{D}$ has a lower probability for the smaller SNR values. Albeit, choosing the \textit{pilot period} by the beam coherence method aimed to have the same average drop in the SNR, but the ML$_{CD}$ is more likely to have lower SNR values in that case due to additional errors caused by estimating the slope variations rather than the AoA estimation errors. 

Fig. \ref{fig:CDFR}, involve the three cases in Fig. \ref{fig:CDFSNR} but in terms of the achievable rate. We can see that for case 1, $P^{}_{out} =0.05$ is satisfied for both ML$_{CD}$, and ML$_{D}$ at the 50\% of the maximum rate. Moreover, the ML$_{CD}$ surpass the ML$_{D}$ roughly by 4 bps/Hz for the same outage without reducing the \textit{pilot period}. For case 2, both ML$_{CD}$, and ML$_{D}$ are nearly the same with minor advantage to the ML$_{CD}$ at the higher rate. Similar to our discussion for Fig. \ref{fig:CDFSNR}, ML$_{D}$ obligated to reduce its \textit{pilot period} by almost 59\% to support $0.95\,\mathcal{R}^{}_{max}$ instead of $0.5\,\mathcal{R}^{}_{max}$ while ML$_{CD}$ only reduce its period by 17\%. Case 3, shows both algorithms can have $P^{}_{out} =0.05$ but only for a very small rate. Still we can notice an advantage to ML$_{D}$ in terms of outage probability since we force both algorithms to drop to the same level which worsen the situation of the ML$_{CD}$ due to additional slope variations errors. From our discussions about Fig. \ref{fig:CDFSNR}, and \ref{fig:CDFR}, we can say that its better to choose the \textit{pilot period} based on the outage probability definition due to:(1) the beam coherence method is not directly related to the operational rate and the outage probability (2) the beam coherence method is not a fair choosing method especially if additional estimation errors exists in the ML$_{CD}$.
\begin{figure}
    \centering
    \includegraphics[scale=0.6]{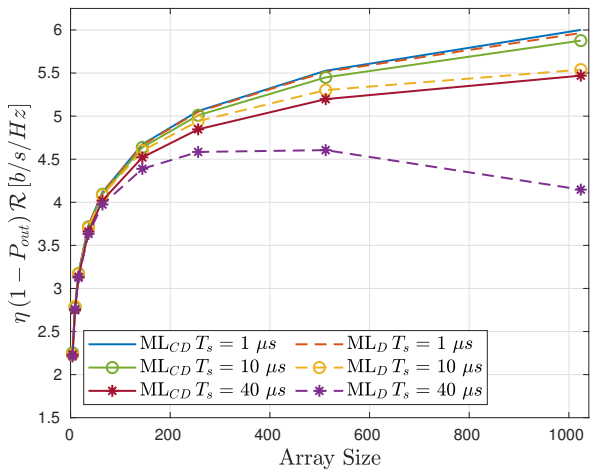}
    \caption{Achievable rate for $\mathcal{R}^{}_{f} {=} 0.5\,\mathcal{R}^{}_{max}$}
    \label{fig:achrate-0.5}
\end{figure}
\begin{figure}
    \centering
    \includegraphics[scale=0.6]{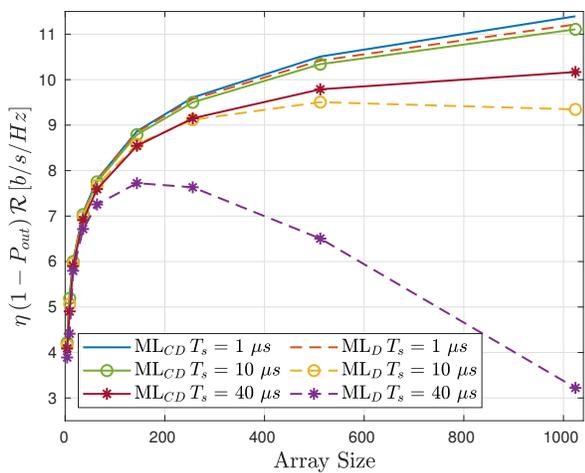}
    \caption{Achievable rate for $\mathcal{R}^{}_{f} {=} 0.95\,\mathcal{R}^{}_{max}$}
    \label{fig:achrate-0.95}
\end{figure}

Up to this point, we have argued the overhead reduction. Now, we measure the performance of the ML$_{CD}$ in terms of the effective rate given by \cref{eqn:reff} when the achievable rate $\mathcal{R}^{}_{a}$ is either the outage rate $(1-P^{}_{out})\mathcal{R}^{}_{f}$ or the average rate $\mathbb{E}[\mathcal{R}]$. First, for the effective outage rate as shown in Fig. \ref{fig:achrate-0.5}, the ML$_{CD}$ outperforms the ML$_{D}$ for all pilot training intervals values. Moreover, we can notice that ML$_{CD}$ is more resilient to discontinuities such that it reduced only by 0.5 bps/Hz for 1024 array element. On contrary, the ML$_{D}$ is more sensitive to pilots disruptions causing a rate drop by 2 bps/Hz for $T^{}_{s} = 40\,\mu s$. 
\begin{figure}
    \centering
    \includegraphics[scale=0.6]{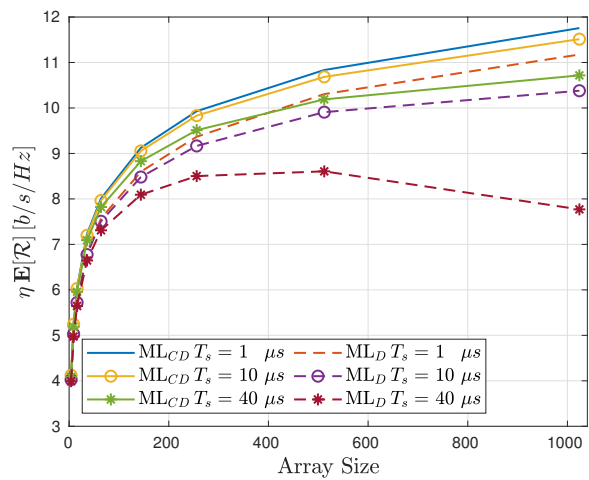}
    \caption{Average achievable rate for $\mathcal{R}^{}_{f} {=} 0.5\,\mathcal{R}^{}_{max}$}
    \label{fig:avgrate0.5}
\end{figure}
\begin{figure}
    \centering
    \includegraphics[scale=0.6]{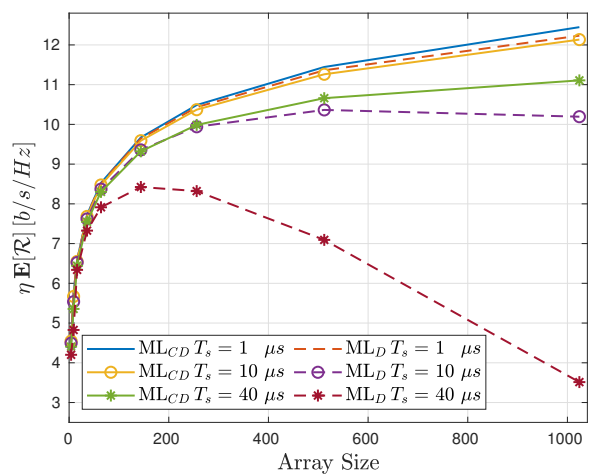}
    \caption{Average achievable rate for $\mathcal{R}^{}_{f} {=} 0.95\,\mathcal{R}^{}_{max}$}
    \label{fig:avgrate0.95}
\end{figure}

Increasing the operational rate to $0.95\,\mathcal{R}^{}_{max}$ increases the pilot overhead reduction which gives the advantage to the ML$_{CD}$ algorithm. This makes ML$_{CD}$ to surpass the ML$_{D}$ by 0.25 bps/Hz for the smallest pilot training interval $T_s = 1\,\mu s$. Moreover, ML$_{CD}$ has a rate drop by nearly 1 bps/Hz for $T^{}_{s}= 40\,\mu s$ while ML$_{D}$ suffers a rate drop up to 8 bps/Hz. Finally, the higher pilot disruptions makes the efficiency slope decreases too fast, and in that case the array gain is not able to compensate. This noticed for the ML$_{D}$ with $T^{}_{s} = 40\,\mu s$, where the rate begins to drastically drop instead of increasing with the array size.

Now, we consider the effective average achievable rate as shown in Fig. \ref{fig:avgrate0.5}, and Fig. \ref{fig:avgrate0.95}. as a verification way of the Continuous-Discrete advantage. Not only the ML$_{CD}$ gain from the overhead reduction or higher efficiency, but also it is more likely to have the higher received SNR than ML$_{D}$. The average rate is more significant in case of $0.5\,\mathcal{R}^{}_{max}$ than $0.95\,\mathcal{R}^{}_{max}$ since in the later case both algorithm are expected to operate near the edge of the maximum rate. This makes the average of both algorithms to be almost the same.

Partially, we have assumed fixed $T^{}_{LR}$ for both approaches. Two more scenarios can be considered when both approaches are assigned the same \textit{pilot period} and that period is designed for (1) the Discrete tracking algorithms and (2) the Continuous-Discrete tracking algorithms. In these two scenarios, Continuous-Discrete tracking will have longer $T^{}_{LR}$ than the Discrete algorithm.
% \begin{figure*}
% \img{WInsite/1.PNG}{Sample route of the UE in urban city}{fig:WIroute}
% \img{WInsite/AoA_NN2_e.eps}{AoA tracking for a practical sample path}{fig:WIAoA}
% % {Angle of Arrival Tracking for a practical sample path}{fig:WIAoA}
% \img{WInsite/SNR_10dB_EKF_ML_UPA_e.eps}{Average SNR for tracking in a practical situation}{fig:WISNR}
% % \vspace{-1.5em}
% \end{figure*}

\begin{figure*}
% \img{Outage2/CDFOut_gamma_SNRpA_8_Q_1000_N_256_F.eps}{CDF of the received SNR}{fig:CDFSNR}
% \img{Outage2/CDFOut_Rate_SNRpA_8_Q_1000_N_256_axis_F.eps}{CDF of the achievable rate}{fig:CDFR}
% \img{Outage2/OutRate_SNRpA_8_Q_1000_A_0.5_axis.eps}{Achievable rate for $\mathcal{R}^{}_{f} {=} 0.5\,\mathcal{R}^{}_{max}$}{fig:achrate-0.5}
% \vspace{0.5em}
% {Achievable Rate for $P^{}_{out} {=} 0.05$ and $\mathcal{R}^{}_{f} {=} 0.5\,\mathcal{R}^{}_{max}$}{fig:achrate-0.5}
% \img{Outage2/OutRate_SNRpA_8_Q_1000_A_0.95_axis.eps}{Achievable rate for $\mathcal{R}^{}_{f} {=} 0.95\,\mathcal{R}^{}_{max}$}{fig:achrate-0.95}
% {Achievable Rate for $P^{}_{out} {=} 0.05$ and $\mathcal{R}^{}_{f} {=} 0.95\,\mathcal{R}^{}_{max}$}{fig:achrate-0.95}
% \img{Outage2/AvgRate_SNRpA_8_Q_1000_Pout_0.05_0.5R_axis2.eps}{Average achievable rate for $\mathcal{R}^{}_{f} {=} 0.5\,\mathcal{R}^{}_{max}$}{fig:avgrate0.5}
% {Average Achievable Rate for $P^{}_{out} {=} 0.05$ and $\mathcal{R}^{}_{f} {=} 0.5\,\mathcal{R}^{}_{max}$}{fig:avgrate0.5}
% \img{Outage2/AvgRate_SNRpA_8_Q_1000_Pout_0.05_0.95R_axis2.eps}{Average achievable rate for $\mathcal{R}^{}_{f} {=} 0.95\,\mathcal{R}^{}_{max}$}{fig:avgrate0.95}
% \vspace{-1.5em}
% {Average Achievable Rate for $P^{}_{out} {=} 0.05$ and $\mathcal{R}^{}_{f} {=} 0.95\,\mathcal{R}^{}_{max}$}{fig:avgrate0.95}
\end{figure*}

\section{Conclusion}\label{sec:conc}
Beam tracking is crucial for maintaining the quality of established links and avoiding the high-cost initial link establishment process. Traditional beam tracking solutions rely on discrete measurement updates that occur at the instances of pilot signal arrival. Such an approach ignores the first-order beam variation information in between channel measurements.
On the contrary, by considering the continuous nature of channel variation over time, a smarter beam tracking solution should exploit the gradual beam variation and thus attempt to continuously and actively adjust the beam directions even when no measurements are available. This can be achieved by deriving the channel's continuous state transition model.

In this paper, we propose ``Continuous-Discrete'' beam tracking framework, which exploits system information, like the first derivative of beam angles, to allow continuous beam updates while still relying on discrete measurements. Our solution requires less frequent pilot symbols (less overhead) while maintaining similar tracking performance to discrete tracking, and it can achieve much better performance if the pilot frequency is kept the same.
The performance is studied under different SNR levels, array sizes, pilot periods, and two different array configurations, ULA and UPA.
Our Continuous-Discrete solution is shown to outperform Discrete tracking algorithms in terms of overhead reduction by up to 87\%. It is also shown that increasing the array size for fixed SNR and pilot period increases the tracking accuracy as long as the pilot period is comparable to the beam coherence time.
    
Another interesting result is shown where larger MIMO arrays do not necessarily lead to improved beam tracking performance. This result is due to (1) the shorter beam coherence time and (2) the higher link disruption probability, which requires lengthy realignment.
Furthermore, in certain situations, it is demonstrated that uniform planar arrays may provide improved beam tracking performance over ULAs of similar size, in terms of average SNR, due to the longer beam coherence time of UPAs.

\appendices

\section{Proof of Theorem \ref{th:TBD}}
\subsection{Discrete Approach:}\label{apx:th1}\color{black}
\begin{proof}
We need to find the period of time at which the power drops by a given ratio. The power ratio is given by \cref{eqn:zeta}, where $P(t) = 1$ is the perfect alignment power gain, and due to misalignment, the power changes over time as given by \cref{eqn:PR}. Now, we replace $\phi(t+T^{}_{b})$, and $\phi(t)$ by $\phi^{}_{k+1}$ and $\phi^{}_k$ respectively, then the power ratio is given by:
\begin{equation}
    \zeta = \displaystyle\frac{1}{N^2}\left|\sum_{m=0}^{N-1}e^{-j\pi m(\cos(\phi^{}_{k+1})-\cos(\phi^{}_{k}))}\right|^2\\
\end{equation}
Since the power is concentrated on the main-lobe, and using \cref{eqn:approx}, then $\zeta$ is simplified to
\begin{equation}\label{eqn:zeta2}
    \zeta \approx e^{-N^2[\cos(\phi^{}_{k+1})-\cos({\phi}^{}_{k})]^2}
\end{equation}
The discrete representation of the AoA model in \cref{eqn:phit} using \textit{pilot period} $T = T^{}_b$ is given by:
\begin{equation}\label{eqn:Dphi}
    \phi^{}_{k+1} = \phi^{}_k + T^{}_b\,\dot{\phi}^{}_k
\end{equation}
Now, we simplify the cosine difference in \cref{eqn:zeta2} using \cref{eqn:Dphi}, and $T^{}_b\,\dot{\phi}^{}_k \ll 1$ as follows:
\begin{equation}
    \begin{aligned}
        \cos(\phi^{}_{k+1}) &= \cos(\phi^{}_k + T^{}_b\,\dot{\phi}^{}_k)\\
        % &= \cos(\phi^{}_k)\cos(T^{}_b\,\dot{\phi}^{}_k) - \sin(\phi^{}_k)\sin(T^{}_b\,\dot{\phi}^{}_k)\\
        &\approx \cos(\phi^{}_{k})-T^{}_b\,\dot{\phi}^{}_k\,\sin(\phi^{}_k)
    \end{aligned}
\end{equation}
Hence, we have $\cos(\phi^{}_{k+1})-\cos(\phi^{}_k)\approx -T^{}_b\,\dot{\phi}^{}_k\,\sin(\phi^{}_k)$, and this implies the following:
\begin{equation}\label{eqn:zetad}
\zeta = e^{-(NT^{}_b\sin(\phi^{}_k))^2\dot{\phi}^2_k}.
\end{equation}
Now, let $T^{}_{LR} = K\,T^{}_b$, and define $\mu^{}_{\zeta} \triangleq
\mathbb{E}\left[\zeta|T^{}_b\,,\phi^{}_k\right]$, since the slope variation of the AoA is following a Gaussian distribution, then after a period of time $T^{}_{LR}$, and $\dot{\phi}^{}_k \sim \mathcal{N}(0,Q\,T^{}_{LR})$

Hence,
\begin{equation}
    \begin{aligned}
        % \mu^{}_{\zeta} &= \int_{-\infty}^{\infty}\displaystyle\frac{1}{\sqrt{2\pi\,QT_{LR}}}\,e^{-T^2_bN^2\sin^2(\phi_k)\dot{\phi}^2_k}\,e^{-\frac{\dot{\phi}^2_k}{2QT_{LR}}}d{\dot{\phi}_k}\\
        \mu^{}_{\zeta} &= \int_{-\infty}^{\infty}\displaystyle\frac{1}{\sqrt{2\pi\,QT^{}_{LR}}}\,e^{-\left(T^{}_b N\sin(\phi^{}_k)\right)^2x^2}\,e^{-\frac{x^2}{2QT^{}_{LR}}}dx\\
        &= \sqrt{\displaystyle\frac{\sigma^2_d}{QT^{}_{LR}}}\int_{-\infty}^{\infty}\displaystyle\frac{1}{\sqrt{2\pi\,\sigma^2_d}}\,e^{-\frac{x^2}{2\sigma^2_d}}dx\\
        &= \sqrt{\displaystyle{\sigma^2_d}\,/\,{QT^{}_{LR}}},\\
    \end{aligned}
\end{equation}
\noindent where $\sigma^{2}_{d} = QT^{}_{LR}/\left(1+2QT^{}_{LR}\left(T^{}_bN\sin(\phi^{}_k)\right)^2\right)$, then:
\begin{equation}
\begin{aligned}
    % \mu^{}_{\zeta} &= \sqrt{\displaystyle\frac{1}{\left(1+2QT^{}_{LR}\left(T^{}_bN\sin(\phi^{}_k)\right)^2\right)}}\\
    \mu^{}_{\zeta} &= \sqrt{\left[{\left(1+2QT^{}_{LR}\left(T^{}_bN\sin(\phi^{}_k)\right)^2\right)}\right]^{-1}}\\
    \implies T^{}_b &= \sqrt{\displaystyle\frac{1/\mu^2_{\zeta}-1}{2\,Q\,T^{}_{LR}\,\sin^2(\phi^{}_k)N^2}}\\
\end{aligned}
\end{equation}
\end{proof}

\subsection{Continuous-Discrete Approach:}\label{apx:th2}
\color{black}
\begin{proof}
Here, we follow similar steps like in \cref{apx:th1} but we are going to replace $\phi(t)$ by $\phi^{}_k + T^{}_L\hat{\dot{\phi}}^{}_{k-1}$ since in a continuous-discrete tracking we continuously update the beamforming in-between pilot symbols. In that case, the cosine difference in \cref{eqn:zeta2} is given by: 
% \begin{equation}
%     \begin{aligned}
%     % \cos(t+T^{}_L) - \cos(t) 
%         \cos(\phi(t+T^{}_L)) - \cos(\phi(t))
%         &= \cos(\phi^{}_k + T^{}_L\dot{\phi}^{}_k) - \cos(\phi^{}_k + T^{}_L\hat{\dot{\phi}}^{}_{k-1})\\
%         &\approx-T^{}_L\sin(\phi^{}_k)\varepsilon^{}_k,\\
%         % &\approx-T_L\sin(\phi_k)\left[\dot{\phi}_k - \hat{\dot{\phi}}_{k-1}\right]\\
%     \end{aligned}
% \end{equation}
\begin{dmath}
    % \begin{aligned}
    % \cos(t+T^{}_L) - \cos(t) 
        \cos(\phi(t+T^{}_L)) - \cos(\phi(t))
        = \cos(\phi^{}_k + T^{}_L\dot{\phi}^{}_k) - \cos(\phi^{}_k + T^{}_L\hat{\dot{\phi}}^{}_{k-1})
        \approx-T^{}_L\sin(\phi^{}_k)\varepsilon^{}_k,
        % &\approx-T_L\sin(\phi_k)\left[\dot{\phi}_k - \hat{\dot{\phi}}_{k-1}\right]\\
    % \end{aligned}
\end{dmath}
where $\varepsilon^{}_k \triangleq \dot{\phi}^{}_k - \hat{\dot{\phi}}^{}_{k-1}$, and by assuming perfect estimate of the previous slope variation, i.e., $\hat{\dot{\phi}}^{}_{k-1} = \dot{\phi}^{}_{k-1}$, then $\varepsilon^{}_k \sim \mathcal{N}(0,\kappa QT^{}_{LR})$. 
The power ratio in that case is given by:
\begin{equation}\label{eqn:zetacd}
\zeta = e^{-(NT^{}_L\sin(\phi^{}_k))^2\varepsilon^2_k}.
\end{equation}
Hence,
\begin{equation}
    \begin{aligned}
        \mu^{}_{\zeta} &= \int_{-\infty}^{\infty}\displaystyle\frac{e^{-\frac{x^2}{2\kappa\,QT^{}_{LR}}}}{\sqrt{2\pi\,\kappa\,QT^{}_{LR}}}\,e^{-\left(T^{}_L N\sin(\phi^{}_k)\right)^2x^2}\,dx\\
        % &= \sqrt{\displaystyle\frac{\sigma^2_{cd}}{QT^{}_{LR}}}\int_{-\infty}^{\infty}\displaystyle\frac{1}{\sqrt{2\pi\,\sigma^2_{cd}}}\,e^{-\frac{x^2}{2\sigma^2_{cd}}}dx
        % = \sqrt{{\sigma^2_{cd}}\,/\,{\kappa\,QT^{}_{LR}}},\\
    \end{aligned}
\end{equation}
\begin{equation}
\begin{aligned}
    % \mu^{}_{\zeta} &= \sqrt{\displaystyle\frac{1}{\left(1+2\alpha\,QT^{}_{LR}\left(T^{}_LN\sin(\phi^{}_k)\right)^2\right)}}\\
    % \mu^{}_{\zeta} &= \sqrt{\left[{\left(1+2\kappa\,QT^{}_{LR}\left(T^{}_LN\sin(\phi^{}_k)\right)^2\right)}\right]^{-1}}\\
    \implies T^{}_L &= \sqrt{\displaystyle\frac{1/\mu^2_{\zeta}-1}{2\,\kappa\,Q\,T^{}_{LR}\,\sin^2(\phi^{}_k)N^2}}\\
\end{aligned}
\end{equation}
\end{proof}
\section{Proof of $\gamma$ Distribution}\label{apx:th3}
\begin{proof}
\subsection*{Discrete Tracking Approach:}
Here, we need to find the PDF and CDF of the instantaneous SNR. From \cref{meas}, the instantaneous received SNR is given by:
\begin{equation}
    \begin{aligned}
        \gamma^{}_{k} &= \rho\,\left|\displaystyle\frac{1}{N}\sum_{m = 0}^{N-1}e^{-j2\pi\Delta m[\cos({\phi^{}_k}) - \cos({\bar{\phi}^{}_k})]}\right|^2\\
        &\approx \rho\,\zeta^{}_{k}
    \end{aligned}
\end{equation}
\noindent and assuming small slope variations during pilot duration, i.e., $T\dot{\phi}^{}_k \ll 1$, for a given value of $\phi^{}_k$, and using \cref{eqn:zetad} then:
\begin{equation}\label{gammad}
    \gamma^{}_{k} \approx \rho\,e^{-N^2T^2\sin^2(\phi^{}_k)\,\dot{\phi}^{2}_k}
\end{equation}
using the Cumulative Distribution Function (CDF) way: 
\begin{equation}
    \begin{aligned}
        F(\gamma^{}_{k}|\phi^{}_k) &= P(\gamma \le \gamma^{}_{k}|\phi^{}_k)\\
        &= P(\rho\,e^{-N^2T^2\sin^2(\phi^{}_k)\,\dot{\phi}^{2}_k} \le \gamma^{}_{k}|\phi^{}_k)\\
        % &= P(e^{-N^2T^2\sin^2(\phi^{}_k)\,\dot{\phi}^{2}_k} \le \frac{\gamma^{}_{k}}{\rho}|\phi^{}_k)\\
        % &= P({-N^2T^2\sin^2(\phi^{}_k)\,\dot{\phi}^{2}_k} \le \log\left({\gamma^{}_{k}}/{\rho}\right)|\phi^{}_k)\\
        &= P\left(\dot{\phi}^{2}_k \ge \frac{\log\left({\rho}/{\gamma^{}_{k}}\right)}{N^2T^2\sin^2(\phi^{}_k)}|\phi^{}_k\right)\\
        &= 1 - P\left(\dot{\phi}^{2}_k \le \frac{\log\left({\rho}/{\gamma^{}_{k}}\right)}{N^2T^2\sin^2(\phi^{}_k)}|\phi^{}_k\right)\\
\end{aligned}
\end{equation}
since $\dot{\phi}^{}_k \sim \mathcal{N}(0,QT^{}_{LR})$, then the squared of $\dot{\phi}^{}_k$ has a Chi-Squared distribution with one degree of freedom. Hence,
\begin{equation}
    \begin{aligned}
    % P\left(\dot{\phi}^{2}_k \le \frac{\log\left({\rho}/{\gamma^{}_{k}}\right)}{N^2T^2\sin^2(\phi_k)}\right) = 1-2\,Q\left(\sqrt{\frac{\log\left({\rho}/{\gamma^{}_{k}}\right)}{N^2T^2\sin^2(\phi_k)QkT}}\right)\\
    P\left(\dot{\phi}^{2}_k \le \dot{\varphi}\right) = 1-2\,\mathbb{Q}\left(\sqrt{\frac{\dot{\varphi}}{QT^{}_{LR}}}\right)\\
    \end{aligned}
\end{equation}
This implies;
\begin{equation}\label{eqn:CDFD}
    F(\gamma^{}_{k}|\phi^{}_k) = 2\,\mathbb{Q}\left(\sqrt{\displaystyle\frac{\log\left({\rho}/{\gamma^{}_{k}}\right)}{N^2T^2\sin^2(\phi^{}_k)QT^{}_{LR}}}\right)
\end{equation}
The Probability Density Function (PDF) is found by taking the derivative of the CDF:
\begin{equation}
    \begin{aligned}
\hspace{-0.94em}f(\gamma^{}_{k}|\phi^{}_k)\hspace{-3px}&= \frac{\partial F(\gamma^{}_{k}|\phi^{}_k)}{\partial \gamma^{}_{k}}\\
    &= 2\frac{\partial }{\partial \gamma^{}_{k}}\hspace{-0.45em}\left(\hspace{-0.3em}\int_{\hspace{-0.5em}\sqrt{\log(\frac{\rho}{\gamma^{}_k})}}^{\infty}\hspace{-0.2em}\frac{e^{-\frac{t^2}{2N^2T^2\sin^2(\phi^{}_k)QT^{}_{LR}}}}{\sqrt{2\pi N^2T^2\sin^2(\phi^{}_k)QT^{}_{LR}}}dt\hspace{-0.3em}\right)\\
    &= {e^{-\frac{\log(\rho/\gamma^{}_k)}{2N^2T^2\sin^2(\phi^{}_k)QT^{}_{LR}}}}\bigg/{\gamma^{}_k\,\sqrt{\log(\rho/\gamma^{}_k)}}
    \end{aligned}\hspace{-10px}
\end{equation}
\subsection*{Continuous-Discrete Tracking Approach:}
Following a similar procedure as in the discrete case, and using \cref{eqn:zetacd} then,
\begin{equation}\label{gammacd}
    \gamma^{}_{k} = \rho\,e^{\displaystyle-N^2T^2\sin^2(\phi^{}_k)\,\varepsilon^2_k}
\end{equation}
Here, we can see the difference between \cref{gammad,gammacd} is replacing $\dot{\phi}^{}_k$ by $\varepsilon^{}_k \sim \mathcal{N}(0,\kappa\,QT^{}_{LR})$, then the CDF is given by:

\begin{equation}\label{eqn:CDFCD}
    \begin{aligned}
    F(\gamma^{}_{k}|\phi^{}_k) &{=} 1- P\left(\varepsilon^2_k \le \displaystyle\frac{\log(\rho/\gamma^{}_k)}{N^2T^2\sin^2(\phi^{}_k)}|\phi^{}_k\right)\\
    &{=} 1-1+2\,\mathbb{Q}\left(\sqrt{\displaystyle\frac{\log(\rho/\gamma^{}_k)}{N^2T^2\sin^2(\phi^{}_k)\kappa\,QT^{}_{LR}}}\right)\\    
    &{=} 2\,\mathbb{Q}\left(\sqrt{\displaystyle\frac{\log(\rho/\gamma^{}_k)}{N^2T^2\sin^2(\phi^{}_k)\kappa\,QT^{}_{LR}}}\right)\\
    \end{aligned}
\end{equation}
Similarly, the PDF is given by:
\begin{equation}
    \begin{aligned}
    f(\gamma^{}_{k}|\phi^{}_k)
    % &= 
    % &= \displaystyle\frac{e^{-\frac{\log(\rho/\gamma^{}_k)}{2N^2T^2\sin^2(\phi^{}_k)\alpha\,QT^{}_{LR}}}}{\gamma^{}_k\,\sqrt{\log(\rho/\gamma^{}_k)}}
    &{=} {e^{-\frac{\log(\rho/\gamma^{}_k)}{2N^2T^2\sin^2(\phi^{}_k)\kappa\,QT^{}_{LR}}}}\,\bigg/\,{\gamma^{}_k\,\sqrt{\log(\rho/\gamma^{}_k)}}
    \end{aligned}
\end{equation}
\end{proof}

\bibliographystyle{IEEEtran}
\bibliography{References}

\begin{IEEEbiography}
[{\includegraphics[width=1in,height=1.25in,clip,keepaspectratio]{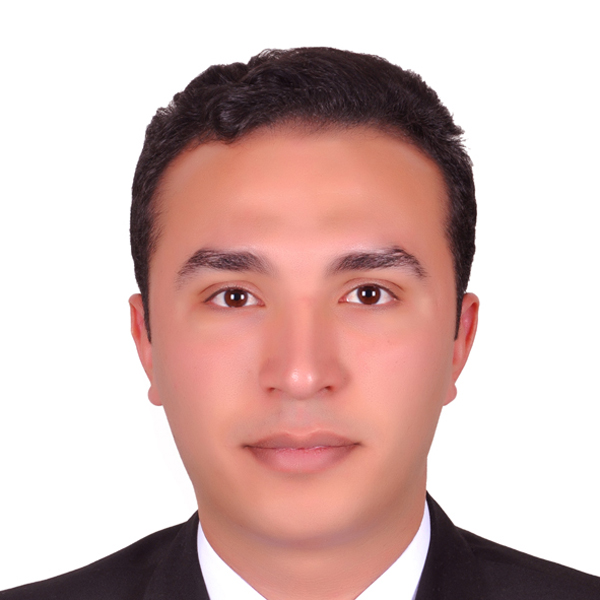}}]
% [{\includegraphics[width=1in,height=1.25in]{Authors/mnaguib.jpg}}]
{Mohamed Naguib} received his B.Sc. (with Highest Honors) and M.Sc. degrees
in electrical engineering from the Military Technical College (MTC), Cairo, Egypt, in 2012 and 2018,
respectively. He received the Ph.D.
degree from the Department of Electrical and Computer Engineering at the Ohio State University. He served as a Teaching and Research Associate with the Department of Communications Engineering at MTC from 2013 to 2020 and is currently a Faculty Member at MTC. His research interests
include wireless communications, information theory, software-defined radio, networking, and cognitive radio networks.
\end{IEEEbiography}

\begin{IEEEbiography}[{\includegraphics[width=1in,height=1.25in,clip,keepaspectratio]{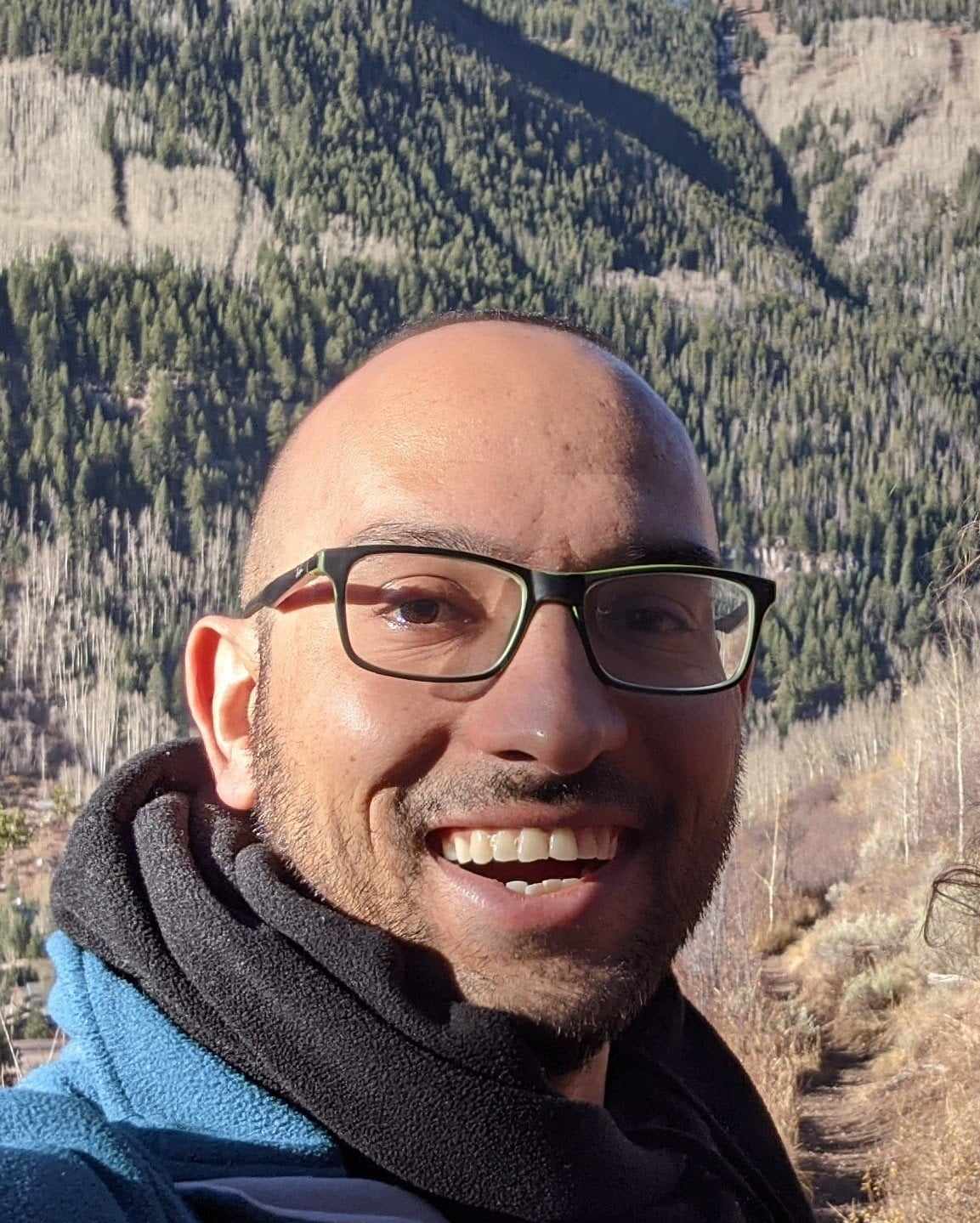}}]{Yahia Shabara} received his Ph.D. degree from the department of Electrical and Computer Engineering at The Ohio State University. Previously he received his B.Sc. degree in Electrical Engineering from Alexandria University, Alexandria, Egypt, in 2012 and the M.Sc. degree in wireless communications from Nile University, Giza, Egypt, in 2015. His research interests include wireless communications, computer networks, machine learning, information theory and network security.
\end{IEEEbiography}
% \vspace{-15em}
\begin{IEEEbiography}[{\includegraphics[width=1in,height=1.25in,clip,keepaspectratio]{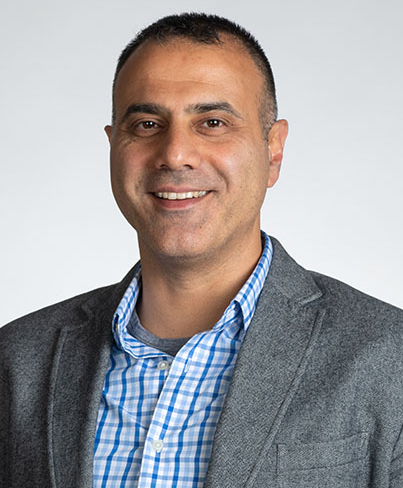}}]{Can Emre Koksal} is the Founder and CEO of Anchor. He is also a Professor of Electrical and Computer Engineering at The Ohio State University. 

Emre received S.M. and Ph.D. degrees from MIT in 1998 and 2003, respectively, in Electrical Engineering and Computer Science. During his studies, he was affiliated with MIT Lincoln Lab., Lab. for Information and Decision Systems, and Computer Science and Artificial Intelligence Labs. He also worked at Sycamore Networks as an intern engineer, prior to their IPO. His areas of expertise include wireless communication, information security, communication networks, and information theory. He holds 8 US and International Patents, licensed and are being commercialized by several companies. He is the author of more than 100 papers, published in top journals and conference proceedings. He has testified to the Congress multiple times on various aspects of cybersecurity.

Emre is the recipient of Columbus Business First – Inventor of the Year Award in 2020, the National Science Foundation CAREER Award in 2011, a finalist of the Bell Labs Prize in 2015, OSU CoE Innovator Award in 2016 and 2020, OSU CoE Lumley Research Award in 2011 and 2017, and the recipient of an HP Labs - Innovation Research Award in 2011. Papers he co-authored received the best paper award in IEEE WiOpt 2018 and the best student paper candidate in ACM MOBICOM 2005. He has served as an Associate Editor for IEEE Transactions on Information Theory, IEEE Transactions on Wireless Communications, and Elsevier Computer Networks.
\end{IEEEbiography}
\end{document}